\documentclass[12pt]{article}
\usepackage{fancyhdr}
\usepackage{multicol}
\usepackage{mathtools}
\usepackage{graphicx}
\graphicspath{ {./images/} }
\usepackage[dvipsnames]{xcolor}
\usepackage{mathrsfs}
\usepackage[T1]{fontenc}
\usepackage{mathpazo}
\usepackage{setspace}
\usepackage{amsfonts}
\usepackage{amssymb}
\usepackage{amsmath}
\usepackage{epsfig}
\usepackage{latexsym}
\usepackage{color}
\usepackage{graphicx}
\usepackage{nicefrac}
\usepackage[latin1]{inputenc}
\usepackage{slashed}
\usepackage{multirow}
\usepackage{comment}
\usepackage{soul}
\usepackage{hyperref}
\usepackage{cite}
\usepackage{datetime}
\usepackage{colortbl}

\usepackage[titletoc]{appendix}

\def\hybrid{\topmargin -20pt    \oddsidemargin 0pt
	\headheight 0pt \headsep 0pt
	\textwidth 6.25in       
	\textheight 9.25in       
	\marginparwidth .875in
	\parskip 5pt plus 1pt   \jot = 1.5ex}

\hybrid

\def\baselinestretch{1.2}

\catcode`\@=11

\def\marginnote#1{}
\newcount\hour
\newcount\minute
\newtoks\amorpm
\hour=\time\divide\hour by60
\minute=\time{\multiply\hour by60 \global\advance\minute by-\hour}
\edef\standardtime{{\ifnum\hour<12 \global\amorpm={am}%
		\else\global\amorpm={pm}\advance\hour by-12 \fi
		\ifnum\hour=0 \hour=12 \fi
		\number\hour:\ifnum\minute<10 0\fi\number\minute\the\amorpm}}
\edef\militarytime{\number\hour:\ifnum\minute<10 0\fi\number\minute}

\def\draftlabel#1{{\@bsphack\if@filesw {\let\thepage\relax
			\xdef\@gtempa{\write\@auxout{\string
					\newlabel{#1}{{\@currentlabel}{\thepage}}}}}\@gtempa
		\if@nobreak \ifvmode\nobreak\fi\fi\fi\@esphack}
	\gdef\@eqnlabel{#1}}
\def\@eqnlabel{}
\def\@vacuum{}
\def\draftmarginnote#1{\marginpar{\raggedright\scriptsize\tt#1}}

\def\draft{\oddsidemargin -.5truein
	\def\@oddfoot{\sl preliminary draft \hfil
		\rm\thepage\hfil\sl\today\quad\militarytime}
	\let\@evenfoot\@oddfoot \overfullrule 3pt
	\let\label=\draftlabel
	\let\marginnote=\draftmarginnote
	\def\@eqnnum{(\theequation)\rlap{\kern\marginparsep\tt\@eqnlabel}%
		\global\let\@eqnlabel\@vacuum}  }

\def\preprint{\twocolumn\sloppy\flushbottom\parindent 2em
	\leftmargini 2em\leftmarginv .5em\leftmarginvi .5em
	\oddsidemargin -.5in    \evensidemargin -.5in
	\columnsep .4in \footheight 0pt
	\textwidth 10.in        \topmargin  -.4in
	\headheight 12pt \topskip .4in
	\textheight 6.9in \footskip 0pt
	\def\@oddhead{\thepage\hfil\addtocounter{page}{1}\thepage}
	\let\@evenhead\@oddhead \def\@oddfoot{} \def\@evenfoot{} }

\def\numberbysection{\@addtoreset{equation}{section}
	\def\theequation{\thesection.\arabic{equation}}}

\def\underline#1{\relax\ifmmode\@@underline#1\else
	$\@@underline{\hbox{#1}}$\relax\fi}

\def\titlepage{\@restonecolfalse\if@twocolumn\@restonecoltrue\onecolumn
	\else \newpage \fi \thispagestyle{empty}\c@page\z@
	\def\thefootnote{\fnsymbol{footnote}} }

\def\endtitlepage{\if@restonecol\twocolumn \else \newpage \fi
	\def\thefootnote{\arabic{footnote}}
	\setcounter{footnote}{0}}  

\catcode`@=12
\relax

\def\figcap{\section*{Figure Captions\markboth
		{FIGURECAPTIONS}{FIGURECAPTIONS}}\list
	{Figure \arabic{enumi}:\hfill}{\settowidth\labelwidth{Figure
			999:}
		\leftmargin\labelwidth
		\advance\leftmargin\labelsep\usecounter{enumi}}}
 \relax
\def\tablecap{\section*{Table Captions\markboth
		{TABLECAPTIONS}{TABLECAPTIONS}}\list
	{Table \arabic{enumi}:\hfill}{\settowidth\labelwidth{Table
			999:}
		\leftmargin\labelwidth
		\advance\leftmargin\labelsep\usecounter{enumi}}}
 \relax
\def\reflist{\section*{References\markboth
		{REFLIST}{REFLIST}}\list
	{[\arabic{enumi}]\hfill}{\settowidth\labelwidth{[999]}
		\leftmargin\labelwidth
		\advance\leftmargin\labelsep\usecounter{enumi}}}
 \relax
\makeatletter
\newcounter{pubctr}
\def\publist{\@ifnextchar[{\@publist}{\@@publist}}
\def\@publist[#1]{\list
	{[\arabic{pubctr}]\hfill}{\settowidth\labelwidth{[999]}
		\leftmargin\labelwidth
		\advance\leftmargin\labelsep
		\@nmbrlisttrue\def\@listctr{pubctr}
		\setcounter{pubctr}{#1}\addtocounter{pubctr}{-1}}}
\def\@@publist{\list
	{[\arabic{pubctr}]\hfill}{\settowidth\labelwidth{[999]}
		\leftmargin\labelwidth
		\advance\leftmargin\labelsep
		\@nmbrlisttrue\def\@listctr{pubctr}}}
 \relax
\makeatother
\newskip\humongous \humongous=0pt plus 1000pt minus 1000pt

\newif\ifdtup

\relax

\def\be{\begin{equation}}
\def\ee{\end{equation}}
\def\ba{\begin{eqnarray}}
\def\ea{\end{eqnarray}}

\def\k{\kappa}

\def\a{\alpha}

\def\b{\beta}

\def\e{\epsilon}

\def\p{\pi}

\def\m{\mu}

\def\l{\lambda}
\def\L{\Lambda}
\def\s{\sigma}
\def\S{\Sigma}

\def\no{\noindent}

\def\IR{\relax{\rm I\kern-.18em R}}

\def \J {{\bar J} }

\def\IR{\relax{\rm I\kern-.18em R}}
\def\IL{\relax{\rm I\kern-.18em L}}

\def\inv{^{\raise.15ex\hbox{${\scriptscriptstyle -}$}\kern-.05em 1}}

\def\tr{{\rm tr}}
\def\Tr{{\rm Tr}}

\begin{document}
	\renewcommand{\theequation}{\thesection.\arabic{equation}}
	\csname @addtoreset\endcsname{equation}{section}
	
	\newcommand{\beq}{\begin{equation}}
	\newcommand{\eeq}[1]{\label{#1}\end{equation}}
	\newcommand{\ber}{\begin{equation}}
	\newcommand{\eer}[1]{\label{#1}\end{equation}}
	\newcommand{\eqn}[1]{(\ref{#1})}
	\begin{titlepage}
		\begin{center}
			

			${}$
			\vskip .2 in

			{\Large\bf Integrable branes in generalized $\l$-deformations}
			
			\vskip 0.4in
			
		 {\bf Georgios P. D. Pappas}
			\vskip 0.1in

			{\em
				Department of Nuclear and Particle Physics,\\
				Faculty of Physics, National and Kapodistrian University of Athens,\\
				15784 Athens, Greece\\
			}

			\vskip 0.1in

			{ \footnotesize{geopappas@phys.uoa.gr}}
			
			
			\vskip .5in
		\end{center}
		
		\centerline{\bf Abstract}
		
		\no
	We search for integrable boundary conditions and their geometric interpretation as $D$-branes, in models constructed as generalized $\l$-deformations of products of group- and coset-spaces. Using the sigma-model approach, we find that all the conformal brane geometries known in the literature for a product of WZW models solve the corresponding boundary conditions, thus persisting as integrable branes along the RG flows of our sigma-models. They consist of the well known $G$-conjugacy classes, twisted $G$-conjugacy classes by a permutation automorphism (permutation branes) and generalized permutation branes. Subsequently, we study the properties of the aforementioned brane geometries, especially of those embedded in the backgrounds interpolating between the UV and IR fixed points.

		\vskip .4in
		\noindent
	\end{titlepage}
	\vfill
	\eject
	
	\newpage
	
	\tableofcontents
	
	\noindent

	\def\baselinestretch{1.2}
	\baselineskip 20 pt
	\noindent
	

	\setcounter{equation}{0}
	
\section{Introduction}
\noindent An extensive amount of literature devoted to the analysis of D-branes, with approaches ranging from algebraic techniques \cite{Ishikawa:2001zu,Gaberdiel:2002qa,Felder:1999ka,Fredenhagen:2001kw} to geometric descriptions \cite{Alekseev:1998mc, Kubota:2001ai,Stanciu:1999id,Stanciu:2000fz,Bachas:2000ik,Elitzur:2001qd,Gawedzki:1999bq,Stanciu:2001vw,Gawedzki:2001ye}, provides us with their complete picture in group spaces $G$ and coset spaces $G/H$. In the first case the D-branes are wrapped around a finite set of allowed $G$-conjugacy classes, denoted with $\mathcal{C}_G$, while in the second one, they are described by the product $\mathcal{C}_G\mathcal{C}_{H}$ of conjugacy classes.

\noindent If we extend the WZW model to a product space $G\times G$ there is an additional class of maximally symmetric branes that can be defined, which include the permutation symmetry between the two manifolds and are known as permutation branes. They have been constructed algebraically in \cite{Recknagel:2002qq} and geometrically in \cite{Figueroa-OFarrill:2000gfl,Sarkissian:2003yw}.

\noindent Since the construction of the permutation branes involves the permutation symmetry between the two manifolds they can be defined only in the case where the two WZW models have the same level $k$. In \cite{Fredenhagen:2005an} the authors suggested a generalization of permutation branes in products of Lie groups with different levels, known as generalized permutation branes (GPB). Unlike the previous ones they are not maximally symmetric, but the symmetry of the diagonally embedded group $G$ in the product space is still conserved. The conformality of these D-branes has been established geometrically in \cite{Fredenhagen:2005an,Fredenhagen:2009hx}. An algebraic construction has been presented in \cite{99}, but only in the case of a product of $N=2$ minimal models with $k_1=1$, $k_2=4$.

\noindent All the brane geometries mentioned above are solutions of boundary conditions in WZW and gauged WZW models, thus they are branes embedded in conformal target spaces. For a general non-linear sigma-model finding consistent boundary conditions and the corresponding brane geometries they describe is both interesting and challenging. Nevertheless, progress can be made in the case where the sigma-model is an integrable model. In this case one can define boundary conditions preserving its integrable structure. These were introduced in \cite{Sklyanin:1988yz} and subsequently applied in several examples of integrable field theories such as the affine Toda field theories \cite{Delius:1998he}, the Green-Schwarz sigma-model \cite{Dekel:2011ja} and the principal chiral model (PCM) \cite{Mann:2006rh}. 

\noindent Another well known class of integrable models are the single $\l$-deformed sigma-models which appear as deformations of a group $G$ or a coset $G/H$ CFT. A construction of these deformations has been done in \cite{Sfetsos:2013wia,Hollowood:2014rla} and is based on the gauging of a WZW field with a principal chiral field. Integrable boundary conditions have been found for these models in \cite{Driezen:2018glg,Driezen:2019ykp}, where in particular the elegant geometric picture of the corresponding integrable branes was presented.

\noindent Inspired by these works, we will search for integrable boundary conditions and their geometric interpretation as $D$-branes, in a specific class of generalized $\l$-deformed models. The first such class is constructed in \cite{Georgiou:2016urf,Georgiou:2017aei,Georgiou:2017oly} and represents the effective action of coupled WZW models, all at the same level $k$, with the characteristic that when the deviation from the conformal point is small they mutually interact via current bilinears forming a closed chain,\footnote{ The currents $J_{i\pm}$ are defined as
	\begin{equation*}
	J_{i+}=\partial_+g_ig_i^{-1},\quad J_{i-}=g_i^{-1}\partial_-g_i.
	\end{equation*}} i.e.
\begin{equation}\label{qqq}
S_{k;\l_1,\dots,\l_N}=\sum_{i=1}^{N}S_{\text{WZW},k}(g_i)-\frac{k}{\pi}\sum_{i=1}^{N}\int d^2\s\,\Tr(J_{i+1+}\l_{i+1}J_{i-})+\mathcal{O}(\l^2)\,,
\end{equation}
where here and in the rest of the paper the index $i$ is defined mod$(N)$. Note that the operators driving the theory from the UV point couple the currents of adjacent copies of the Lie Group $G$. A Hamiltonian analysis of the model revealed it's canonical equivalence to $N$ independent single $\l$-deformed models with couplings $\l_1,\dots, \l_N$. Thus the RG flow equations of 
each of the coupling matrices $\l_i$ are the same  as that of a single $\l$-deformed model \cite{Georgiou:2017oly}. The second class, for which we will search for integrable branes, was introduced in \cite{Georgiou:2017jfi} and appears as a double deformation of a product of WZW models defined at different levels. Compared to \eqref{qqq}, the effective action for the $N=2$ case, is slightly modified and takes the form 
\begin{equation}\label{qqqq}
S_{k_1,k_2;\l_1,\l_2}=S_{k_1}(g_1)+S_{k_2}(g_2)-\frac{\sqrt{k_1k_2}}{\pi}\int d^2\s\Tr(J_{1+}\l_1J_{2-}+J_{2+}\l_2J_{1-})+\mathcal{O}(\l^2)
\end{equation}
It was shown in \cite{Georgiou:2017jfi} that due to the different levels the RG flow acquires a fixed point in the IR, in which the CFT is given as a product of current- and coset-type symmetries (see \eqref{312}). Generalizations of \eqref{qqqq} in the case of an arbitrary number of WZW models has been done in \cite{Georgiou:2020eoo}, where a detailed analysis of the IR CFTs showed that although they are characterized by an asymmetry between their holomorphic (right) and antiholomorphic (left) symmetry algebras the left and right sectors posses the same central charge, i.e. $c_L=c_R$. Finally, the third class of models was constructed in \cite{Sfetsos:2014cea} and further studied in \cite{Sfetsos:2017sep} and is given as a deformation of a diagonal coset space CFT $G_{k_1}\times G_{k_2}/G_{k_1+k_2}$. In these works it has been shown that the operator driving the theory from the UV point is a parafermion bilinear. Additionally the theory smoothly flows to an IR unitary CFT given as $G_{k_1-k_2}\times G_{k_2}/G_{k_1}$ \cite{Sfetsos:2017sep}.

\noindent Since the models presented above appear as integrable deformations of product spaces, they have richer mathematical and physical structures compared to the $\l$-deformed cases, a fact that is expected to be reflected in the variety of integrable branes that can be defined in these theories. Specifically, we will see that these models admit as integrable configurations all the consistent brane geometries presented above for the WZW model.

\noindent The paper is organized as follows: In section \ref{Ibc}, we will apply the method of finding integrable boundary conditions in the sigma-models of our interest. In section \ref{Ibg}, we will determine the integrable brane configurations, i.e. the brane geometry and its gauge invariant two-form, corresponding to the boundary conditions found in section \ref{Ibc}. On the contrary to \cite{Driezen:2018glg} and \cite{Driezen:2019ykp}, the task of determining the configurations directly from the integrable conditions is more complicated due to the boundary equations being more involved. Towards this, we will use the sigma-model approach, e.g. \cite{Stanciu:2000fz} \cite{Elitzur:2001qd} \cite{Sark2} \cite{Gawedzki:2001rm}. Having determined the integrable brane geometries, in section \ref{cftlimits} we will investigate the symmetries they preserve in the two conformal points of the RG flows described in \cite{Georgiou:2017jfi} and \cite{Sfetsos:2017sep}. In section \ref{gpb}, we will study the generalized permutation branes embedded in the aforementioned RG flows  \cite{Georgiou:2017jfi} and \cite{Sfetsos:2017sep}. For the particular case with one of the couplings set to zero, where the model interpolates between two current algebra CFTs \cite{Georgiou:2017jfi}, and for $G=SU(2)$ we will present explicitly the induced background fields (metric and $H$-field) on the lowest dimensional GPB. Finally, in Appendix \ref{A} we apply the sigma-model approach in the group and coset space $\l$-models and in the Appendix \eqref{B''} we present the detailed computations of the boundary conditions to which our branes correspond.

\section{Integrable boundary conditions}\label{Ibc}
In this section we will closely follow \cite{Dekel:2011ja,Mann:2006rh,Driezen:2018glg,Driezen:2019ykp} for obtaining integrability preserving boundary conditions for a class of generalized $\l$-deformed models.  
\subsection{The isotropic deformation at equal levels}\label{ide}
Let us consider the $\l$-deformation of two WZW models, at the same level $k$. This model, referred from now on as model (I), is described through the action \cite{Georgiou:2016urf}
\begin{equation}\label{16'}
\begin{split}
&S_{k;\l_1,\l_2}=S_{\text{WZW},k}(g_1)+S_{\text{WZW},k}(g_2)-\\
&-\frac{k}{\pi}\int_{\S}d^2\s\Tr \left[\begin{pmatrix}
J_{1+}&J_{2+}
\end{pmatrix}\begin{pmatrix}
\l_1\l_2\mathcal{O}_{21}D_2^T&\l_1\mathcal{O}_{21}\\
\l_2\mathcal{O}_{12}& \l_1\l_2\mathcal{O}_{12}D_1^T
\end{pmatrix}\begin{pmatrix}
J_{1-}\\J_{2-}
\end{pmatrix}\right],
\end{split}
\end{equation}
where we assumed isotropic couplings $\l_1, \l_2$ and we introduced the operator
\begin{equation}
\mathcal{O}_{12}=(\mathbb{1}-\l_1\l_2D^T_1D^T_2)^{-1}\,,
\end{equation} 
which is given in terms of the adjoint operator, $D_i(X)=g_i(X)g_i^{-1}$ for $i=1,2$ and $X\in\mathfrak{g}$. Its equations of motion can be encoded in two independent sets of first order differential equations
\begin{equation}\label{eom}
\partial_{\pm}A_{i\mp}=\pm\frac{1}{1+\l_i}[A_{i+},A_{i-}]\,,\quad i=1,2\,, 
\end{equation}
where the algebra valued fields $A_{1\pm}, A_{2\pm}\in\mathfrak{g}$ are given in terms of the group elements $g_1, g_2\in G$ as
\begin{equation}\label{1a}
\begin{split}
&A_{1+}=\l_1(\mathbb{1}-\l_1\l_2D_1D_2)^{-1}(J_{1+}+\l_2D_1J_{2+})\,,\\
&A_{1-}=-\l_1(\mathbb{1}-\l_1\l_2D_2^TD_1^T)^{-1}(J_{2-}+\l_2D_2^TJ_{1-})\,,\\
&A_{2+}=\l_2(\mathbb{1}-\l_1\l_2D_2D_1)^{-1}(J_{2+}+\l_1D_2J_{1+})\,,\\
&A_{2-}=-\l_2(\mathbb{1}-\l_1\l_2D_1^TD_2^T)^{-1}(J_{1-}+\l_1D_1^TJ_{2-})\,,
\end{split}
\end{equation}
and its energy-momentum tensor takes the simple form
\begin{equation}\label{25}
T_{\pm\pm}=k\sum_{i=1}^{2}\frac{1-\l_i^2}{\l_i^2}\Tr(A_{i\pm},A_{i\pm})\,.
\end{equation}
Integrability of the model utilizes the fact that \eqref{eom} can be put in the form of the zero curvature condition \cite{Georgiou:2016urf}
\begin{equation}\label{26}
\partial_+\mathcal{L}_{-}(z)-\partial_-\mathcal{L}_{+}(z)=[\mathcal{L}_{+}(z),\mathcal{L}_{-}(z)],\quad z\in\mathbb{C}\,,
\end{equation}
for the Lax matrices 
\begin{equation}\label{27}
\mathcal{L}_{i\pm}(z_i)=\frac{2z_i}{z_i\mp 1}\frac{1}{1+\l_i}A_{i\pm}\,,\quad  z_i\in \mathbb{C}\,,\quad i=1,2\,.
\end{equation}
Thus, one can construct two transport matrices
\begin{equation}\label{2}
\begin{aligned}
T_i(b,a;z_i)=P\exp\left(-\int_{a}^{b}d\s L_{i\s}(\tau,\s,z_i)\right)\,,\quad L_{i\s}=\frac{1}{2}(L_{i+}-L_{i-})\,,\quad i=1,2\,,
\end{aligned}
\end{equation}
which in the case of a closed string, generate two independent infinite sets of conserved charges. 

\noindent Keeping the discussion general, the integrable structure of a sigma-model defined on an open string, i.e. $\s\in[0,\pi]$, is generically broken, since the underlying monodromy matrix, denoted with $T(z)$, could cease to be conserved. As has been argued in the literature, see e.g. \cite{Delius:1998he,Dekel:2011ja,Mann:2006rh}, in this case the correct object to use as a generating function for integrals of motion, reads
\begin{equation}\label{3}
T_{b}(z)=T^\Omega_R(2\pi,\pi;z)T(\pi,0;z)\,,
\end{equation}
and is known as the boundary monodromy matrix. This involves an integral from one endpoint of the string to the other and then back into the opposite direction. The matrix $T^{\Omega}_R$ is defined in the region $\s_R\in[\pi,2\pi]$ and is constructed from the reflected values of the Lax pair, $R: \mathcal{L}(\s)\to\mathcal{L}^R(\s)$. The superscript $\Omega$ denotes the possibility of including the action of a constant algebra automorphism. Being specific
\begin{equation}\label{210}
T^{\Omega}_R(b,a;z)=P\exp\left(-\int_{a}^bd\s\,\Omega L^R_{\s}(\tau,\s;z)\right)\,.
\end{equation}
Then imposing integrability, i.e. demanding that 
\begin{equation}\label{4}
\partial_{\tau}T_{b}(z)=[T_{b}(z),N]\,,
\end{equation}
for some matrix $N(z)$ one finds appropriate boundary conditions such that \eqref{4} is true. In fact, one may construct more involved boundary matrices as long as their time derivative can be put in the form \eqref{4}, (see e.g. (3.19) in \cite{Driezen:2018glg}).

\noindent Returning to our case we will see that two different definitions of $R$, will lead to integrable boundary conditions which will prove to describe two distinct brane configurations.

\noindent To proceed let us consider the following action of the reflection 
\begin{equation}\label{481}
R:\s\to 2\pi-\s,\quad g_i\to g_{i+1}^{-1}\,,\quad i=1,2\,,
\end{equation}
Under \eqref{481} the WZW currents transform as
 \begin{equation}\label{et''}
 J_{i\pm}(\s)\to J_{iR\pm}(\s)=-J_{i+1\mp}(2\pi-\s)\,,\quad i=1,2\,,
 \end{equation}
 which, with the aid of \eqref{1a}, lead to
 \begin{equation}\label{214}
 A_{i\pm}(\s)\to A_{iR\pm}(\s)=A_{i\mp}(2\pi-\s)\,,\quad i=1,2\,.
 \end{equation}
 Using \eqref{210} and \eqref{214} one can easily show that the reflected transport matrices \eqref{2} satisfy the relation 
 \begin{equation}\label{krop}
 T^{\Omega_i}_{iR}(2\pi,\pi;z)=T^{\Omega_i}_i(0,\pi;-z),\quad i=1,2\,,
 \end{equation} 
 with $\Omega_i$ an automorphism acting on $\mathfrak{g}_1\oplus\mathfrak{g}_2$ and the boundary monodromy matrices \eqref{3} are given as
 \begin{equation}\label{215}
T_{ib}=T^{\Omega_i}_{i}(0,\pi;-z)T_i(\pi,0;z)\,,\quad i=1,2\,.
 \end{equation}
Differentiating \eqref{215} leads to\footnote{To derive \eqref{218} we used the identity
 	\begin{equation}\label{216}
 	\partial_{\tau}T^{\Omega}(b,a;z)=T^{\Omega}(b,a;z)L^{\Omega}_{\tau}(a;z)-L^{\Omega}_{\tau}(b;z)T^{\Omega}(b,a;z)
 	\end{equation}
 	where for simplicity we defined $L^{\Omega}(z)=\Omega L(z)$.}
 \begin{equation}\label{218}
 \begin{split}
 \partial_{\tau}T_{ib}=&[T_{i}^{\Omega_i}(0,\pi;-z)\mathcal{L}^{\Omega_i}_{i\tau}(\pi;-z)-\mathcal{L}^{\Omega_i}_{i\tau}(0;-z)T_{i+1}^{\Omega_i}(0,\pi;-z)]T_i(\pi,0;z)\\
 &+T^{\Omega_i}_{i}(0,\pi;-z)[T_i(\pi,0;z)\mathcal{L}_{i\tau}(0;z)-\mathcal{L}_{i\tau}(\pi;z)T_{i}(\pi,0;z)]\,,\quad i=1,2\,.
 \end{split}
 \end{equation}
 It is a matter of a simple algebra to show that \eqref{218} can be written in the form \eqref{4} for the matrices $N_{i}(z)=\mathcal{L}_{i\tau}(z)$ and the boundary conditions
 \begin{equation}\label{4'}
 \mathcal{L}_{i\tau}(z)|_{\partial\S}=\Omega_i\mathcal{L}_{i\tau}(-z)|_{\partial\S}\,,\quad i=1,2\,,
 \end{equation}
 in both string ends. Substituting in \eqref{4'} the form of the time component of the Lax\footnote{The time components of the Lax pairs are given as
 	\begin{equation}\label{220}
\mathcal{L}_{i\tau}=\frac{z}{z^2-1}\frac{1}{1+\l_i}\left((z+1)A_{i+}+(z-1)A_{i-}\right),\quad i=1,2
\end{equation}} and comparing terms of the same power in $z$ we find that
 \begin{equation}\label{487}
 A_{1+}|_{\partial\S}=\Omega_1 A_{1-}|_{\partial\S}\,,\quad A_{2+}|_{\partial\S}=\Omega_2 A_{2-}|_{\partial\S}
 \end{equation}
As in \cite{Driezen:2018glg}, for consistency reasons the automorphisms $\Omega_i$ must be constant involutive matrices, i.e. $\Omega^2_i=\mathbb{1}$. Demanding additionally that the conditions \eqref{487} lead to \footnote{For a general field theory the condition of the absence of momentum flow across the boundary, translates to the boundary conditions $T_{\tau\s}|_{\partial\S}=0\to T_{++}-T_{--}|_{\partial\S}=0$, known as conformal boundary conditions. In the case where the theory is conformal they preserve one copy of the Virassoro algebra.}
\begin{equation}\label{222}
T_{++}|_{\partial\S}=T_{--}|_{\partial\S}\,,
\end{equation}
and using \eqref{25}, we find that $\Omega_i$ must preserve the trace in the $\mathfrak{g}_1\oplus\mathfrak{g}_2$ algebra, i.e. $\Omega_i^T\Omega_i=\mathbb{1}$, $i=1,2$.

\noindent Let us now define the reflection operator to be
 \begin{equation}\label{488}
R:\s\to 2\pi-\s,\quad g_{i}\to g_i^{-1},\quad \l_i\to \l_{i+1}\,.
\end{equation}
In this case the fields $A_{i\pm}$ transform as
\begin{equation}\label{225}
A_{i\pm}(\s)\to A^R_{i\pm}(\s)=A_{i+1\mp}(2\pi-\s)\,,\quad i=1,2
\end{equation}
and the reflected transport matrix satisfies the relation
\begin{equation}\label{226}
T^{\Omega_i}_{iR}(2\pi,\pi;z)=T^{\Omega_i}_{i+1}(0,\pi;-z)\,,\quad i=1,2\,,
\end{equation}
where we remind to the reader that the index $i$ is defined mod$(2)$. Following the same steps as before we end up with
\begin{equation}\label{227}
\mathcal{L}_{i\tau}(z)|_{\partial\S}=\Omega_i\mathcal{L}_{i+1\tau}(-z)|_{\partial\S}\,,\quad i=1,2\,.
\end{equation}
For consistency reasons \eqref{227} leads to $\Omega_1=\Omega_2^{-1}=\Omega$, where $\Omega$ in comparison with the previous case (see below \eqref{222}), does not need to be involutive. Substituting now \eqref{220} in \eqref{227} we find the boundary conditions
\begin{equation}\label{227'}
\begin{split}
\frac{1}{1+\l_1}A_{1+}|_{\partial\S}=\frac{1}{1+\l_2}\Omega A_{2-}|_{\partial\S}\,,\quad
\frac{1}{1+\l_2}A_{2+}|_{\partial\S}=\frac{1}{1+\l_1}\Omega^{-1}A_{1-}|_{\partial\S}\,.
\end{split}
\end{equation}
Further, demanding that \eqref{227'} are conformal boundary conditions, i.e. satisfying \eqref{222}, we find that $\Omega$ is metric preserving and additionally that the couplings are equal $\l_1=\l_2$. Thus we find the integrable boundary conditions
 \begin{equation}\label{228}
A_{1+}|_{\partial\S}=\Omega A_{2-}|_{\partial\S}\,,\quad A_{2+}|_{\partial\S}=\Omega^{-1} A_{1-}|_{\partial\S}\,.
\end{equation}    
For the reader's convenience we gather the results of the model (I). It admits two sets of integrable boundary conditions, namely
\begin{equation}\label{229}
\begin{pmatrix}
A_{1+}\\
A_{2+}
\end{pmatrix}_{\partial\S}=\begin{pmatrix}
\Omega_1&0\\
0&\Omega_2
\end{pmatrix}\begin{pmatrix}
A_{1-}\\
A_{2-}
\end{pmatrix}_{\partial\S}\,,
\end{equation}
and
\begin{equation}\label{230'}
\begin{pmatrix}
A_{1+}\\
A_{2+}
\end{pmatrix}_{\partial\S}=\begin{pmatrix}
0&\Omega\\
\Omega^{-1}&0
\end{pmatrix}\begin{pmatrix}
A_{1-}\\
A_{2-}
\end{pmatrix}_{\partial\S}\,.
\end{equation}
The former set of boundary conditions do not require equality of the couplings, while in the latter, one has to impose them equal. Note that this constraint does not contradict with their beta functions \cite{Georgiou:2017aei}.

\noindent Already at this level, we notice that in the UV limit  the model (I) describes a product of WZW models $G_k\times G_k$ and the integrable boundary conditions \eqref{229}, \eqref{230'}, for trivial acting automorphisms, reduce to the well known maximally symmetric boundary conditions
\begin{equation}\label{231''}
J_{1+}|_{\partial\S}=-J_{2-}|_{\partial\S}\,,\quad J_{2+}|_{\partial\S}=-J_{1-}|_{\partial\S}\,,
\end{equation}
and
\begin{equation}\label{232}
J_{1+}|_{\partial\S}=-J_{1-}|_{\partial\S}\,,\quad J_{2+}|_{\partial\S}=-J_{2-}|_{\partial\S}\,,
\end{equation}
respectively. Equation \eqref{232} describes branes wrapping around a product of conjugacy classes $\mathcal{C}_{f_1f_2}=\mathcal{C}_{f_1}\times \mathcal{C}_{f_2}$, with $\mathcal{C}_{f_i}=\{hf_ih_i^{-1}|\,\forall h\in G\}$, while \eqref{231''} describe conjugacy classes twisted by a permutation automorphism (permutation branes) \cite{Figueroa-OFarrill:2000gfl,Sarkissian:2003yw}, which we will denote as ${\mathcal{C}^{\pi}}_{f_1f_2}={\mathcal{C}^{\pi}}_{f_1}\times\big({\mathcal{C}^{\pi}}_{f^{-1}_2}\big)^{-1}$, with $\pi(\mathfrak{g}_1,\mathfrak{g}_2)=(\mathfrak{g}_2,\mathfrak{g}_1)$. \footnote{Here ${\mathcal{C}^{\pi}}_{f_i}=\{hf_i\Pi(h_i^{-1})|\,\forall h\in G\}$ and $\Pi$ is defined as $\Pi(h)=\exp(\pi(X))=\exp(X^A\pi(T^A))$.} In later sections, we will realize \eqref{229}, \eqref{230'} geometrically and we will see that the above picture of the maximally symmetric branes survive the deformation, as also shown in \cite{Driezen:2018glg} for the case of the single $\l$-model.         
 
\subsection{The isotropic deformation at unequal levels}
The $\l$-deformation of two WZW models defined at different levels $k_1, k_2$, to be referred from now on as model (II), is described by the action \cite{Georgiou:2017jfi}  
\begin{equation}\label{30a'}
\begin{split}
&S_{k_1,k_2;\l_1,\l_2}=S_{\text{WZW},k_1}(g_1)+S_{\text{WZW},k_2}(g_2)-\\
&-\frac{1}{\pi}\int_{\S}d^2\s\Tr \left[\begin{pmatrix}
J_{1+}&J_{2+}
\end{pmatrix}\begin{pmatrix}
k_1\l_1\l_2\mathcal{O}_{21}D_2^T&k_2\l_0\l_1\mathcal{O}_{21}\\
k_1\l_0^{-1}\l_2\mathcal{O}_{12}& k_2\l_1\l_2\mathcal{O}_{12} D_1^T
\end{pmatrix}\begin{pmatrix}
J_{1-}\\J_{2-}
\end{pmatrix}\right].
\end{split}
\end{equation}
As has been analyzed in \cite{Georgiou:2017jfi, Georgiou:2020eoo}, the difference in the levels results to an RG flow which acquires a new fixed point in the IR for specific values of the couplings, $(\l_1,\l_2)=(\l_0,\l_0)$ with $\l_0=\sqrt{k_1/k_2}$. This level inequality leads to several interesting features for the integrable branes embedded in \eqref{30a'}. These comprise its surviving symmetries, their quantization condition and the consistent definition of the so-called generalized permutation branes in the RG fixed points \cite{Fredenhagen:2005an}. 

\noindent As before its equations of motion can be recast in the form of a zero curvature condition \eqref{26} for the Lax matrices
\begin{equation}\label{234}
\begin{split}
\mathcal{L}_{1\pm}=\frac{2z}{z^2-1}\frac{1-(\l_0)^{\mp 1}\l_1}{1-\l_1^2}A_{1\pm}\,,\quad
\mathcal{L}_{2\pm}=\frac{2z}{z^2-1}\frac{1-(\l_0)^{\pm 1}\l_2}{1-\l_2^2}A_{2\pm}\,,
\end{split}
\end{equation}
with
\begin{equation}\label{235''}
\begin{split}
&A_{1+}=\l_1(\mathbb{1}-\l_1\l_2D_1D_2)^{-1}(\l_0J_{1+}+\l_2D_1J_{2+})\,,\\
&A_{1-}=-\l_1(\mathbb{1}-\l_1\l_2D_2^TD_1^T)^{-1}(\l_0^{-1}J_{2-}+\l_2D_2^TJ_{1-})\,,\\
&A_{2+}=\l_2(\mathbb{1}-\l_1\l_2D_2D_1)^{-1}(\l_0^{-1}J_{2+}+\l_1D_2J_{1+})\,,\\
&A_{2-}=-\l_2(\mathbb{1}-\l_1\l_2D_1^TD_2^T)^{-1}(\l_0J_{1-}+\l_1D_1^TJ_{2-})
\end{split}
\end{equation}
and its energy-momentum tensor reads
\begin{equation}\label{236}
\begin{split}
&T_{++}=k_2\frac{1-\l_1^2}{\l_1^2}\Tr(A_{1+},A_{1+})+k_1\frac{1-\l_2^2}{\l_2^2}\Tr(A_{2+},A_{2+})\,,\\
&T_{--}=k_1\frac{1-\l_1^2}{\l_1^2}\Tr(A_{1-},A_{1-})+k_2\frac{1-\l_2^2}{\l_2^2}\Tr(A_{2-},A_{2-})\,.
\end{split}
\end{equation}
If we define the reflection transformation as 
\begin{equation}\label{tra'}
R:\s\to 2\pi-\s\,,\quad g_i\to g_{i+1}^{-1}\,,\quad\l_0\to\l_0^{-1}\,,\quad i=1,2\,,
\end{equation}
and proceed along the same lines as before we arrive at the boundary conditions
 \begin{equation}\label{237}
 \begin{split}
 &\frac{1-\l_0^{-1}\l_1}{1-\l_1^2}A_{1+}|_{\partial\S}=\frac{1-\l_0\l_1}{1-\l_1^2}\Omega_1A_{1-}|_{\partial\S}\,,\\
 &\frac{1-\l_0\l_2}{1-\l_2^2}A_{2+}|_{\partial\S}=\frac{1-\l_0^{-1}\l_2}{1-\l_2^2}\Omega_2A_{2-}|_{\partial\S}\,.
 \end{split}
\end{equation}
If we instead define that
\begin{equation}\label{238}
R:\s\to 2\pi-\s\,,\quad g_i\to g_{i}^{-1}\,,\quad \l_{i}\to\l_{i+1}\,,\quad i=1,2\,,
\end{equation}
we find the boundary conditions
 \begin{equation}\label{239}
 \begin{split}
&\frac{1-\l_0^{-1}\l_1}{1-\l_1^2}A_{1+}|_{\partial\S}=\frac{1-\l_0^{-1}\l_2}{1-\l_2^2}\Omega A_{2-}|_{\partial\S}\,,\\  &\frac{1-\l_0\l_2}{1-\l_2^2}A_{2+}|_{\partial\S}=\frac{1-\l_0\l_1}{1-\l_1^2}\Omega A_{1-}|_{\partial\S}\,,
\end{split}
\end{equation}
where, as in the case of equal levels, $\Omega$ does not need to be involutive.
 
\noindent Using \eqref{236} and demanding that \eqref{239} satisfy \eqref{222}, we find that $\Omega$ preserves the trace $k_1\Tr(,)+k_2\Tr(,)$ and that  $\l_1=\l_2=\l$ along the whole deformation line.\footnote{As in the case of equal levels the restriction on the running of the couplings is compatible with their beta fuction. Additionally, one can see that in the UV limit where model (II) describes a $G_{k_1}\times G_{k_2}$ CFT, the boundary conditions \eqref{239} describe the product $\mathcal{C}_{f_1f_2}$.} Turning our attention to the other set of boundary conditions \eqref{237} we see that they do not satisfy \eqref{222} \footnote{Unless $\l_0=1$ which is the case already studied} and they do not have a geometric realization in terms of Dirichlet and Neumann conditions. However they comprise consistent integrable boundary conditions.

\noindent Thus model (II) admits one kind of integrable boundary conditions compatible with the vanishing of the momentum flow through the boundary  
\begin{equation}\label{10}
\begin{split}
\begin{pmatrix}
A_{1+}\\
A_{2+}
\end{pmatrix}_{\partial\S}=\begin{pmatrix}
0&\Omega\\
\Omega^{-1}&0
\end{pmatrix}\begin{pmatrix}
A_{1-}\\
A_{2-}
\end{pmatrix}_{\partial\S},\quad \l_1=\l_2\,. 
\end{split}
\end{equation}
\subsection{The isotropic deformation of $G_{k_1}\times G_{k_2}/G_{k_1+k_2}$ space}
As a last example we consider the isotropic $\l$-deformed $G_{k_1}\times G_{k_2}/G_{k_1+k_2}$ coset space \cite{Sfetsos:2017sep}. As it has been shown it has a $G$-gauge invariance given as
\begin{equation}\label{118}
(g_1,g_2)\sim(Lg_1L^{-1},Lg_2L^{-1}),\quad L(\s_+,\s_-)\in G
\end{equation}
and it flows to a coset space CFT in the IR. Its action reads as
 \begin{equation}\label{231'}
\begin{split}
S&_{k_1,k_2,\l}=S_{WZW,k_1}(g_1)+S_{WZW,k_2}(g_2)-\\
&-\frac{k_1}{\pi}\int_{\S}\Tr\left[J_{1+}\L_{12}^{-T}\left((1-\l)(s_1J_{1-}+s_2J_{2-})-4s_1s_2\l(D_2^T-1)J_{1-}\right)\right]\\
&-\frac{k_2}{\pi}\int_{\S}\Tr\left[J_{2+}\L_{21}^{-T}\left((1-\l)(s_1J_{1-}+s_2J_{2-})-4s_1s_2\l(D_1^T-1)J_{2-}\right)\right]\,,
\end{split}
\end{equation}
where $s_i=k_i/k$, $i=1,2$ with $k=k_1+k_2$. It has been proven to be integrable as it admits the flat Lax connection
\begin{equation}\label{11}
L_{\pm}=\mathcal{A}_{\pm}+(a+z^{\pm 1}\sqrt{a^2+\b})\mathcal{B}_{\pm}\,,\quad z\in\mathbb{C}\,,
\end{equation} 
with the coefficients given by
\begin{equation}\label{12}
\begin{split}
\a=-\frac{(s_1-s_2)(1-\l)}{1-\l(1-8s_1s_2)}\,,\quad
\b=\frac{1+\l-2\l^2(1-4s_1s_2)}{\l(1-\l(1-8s_1s_2))}
\end{split}
\end{equation}
and
\begin{equation}
\mathcal{A}_{\pm}=\frac{1}{2}(A_{1\pm}+A_{2\pm})\,,\quad \mathcal{B}_{\pm}=\frac{1}{2}(A_{1\pm}-A_{2\pm})\,.
\end{equation}
The gauge fields $A_{i\pm}$ are given as
\begin{equation}\label{39a}
\begin{split}
&A_{1+}=\L^{-1}_{21}((1-\l)(s_1J_{1+}+s_2J_{1+})-4s_1s_2\l(D_2-1)J_{1+})\,,\\
&A_{2+}=\L^{-1}_{12}((1-\l)(s_1J_{1+}+s_2J_{2+})-4s_1s_2\l(D_1-1)J_{2+})\,,\\
&A_{1-}=-\L^{-T}_{12}((1-\l)(s_1J_{1-}+s_2J_{2-})-4s_1s_2\l(D_2^T-1)J_{1-})\,,\\
&A_{2-}=-\L^{-T}_{21}((1-\l)(s_1J_{1-}+s_2J_{2-})-4s_1s_2\l(D_1^T-1)J_{2-})\,,
\end{split}
\end{equation}
with
\begin{equation}\label{35a}
\begin{split}
&\L_{12}=4\l s_1s_2(D_1-1)(D_2-1)+(\l-1)(s_1D_1+s_2D_2-1)\,.
\end{split}
\end{equation}
As before we built the boundary monodromy matrix
\begin{equation}\label{er}
T_b(z)=T_R(2\pi,\pi;z)T(\pi,0;z)\,,
\end{equation} 
where we considered $\Omega=\mathbb{1}$ and $T_R(2\pi,\pi;z)$ is constructed from the Lax \eqref{11} under the reflection $\s\to 2\pi-\s$ so that
\begin{equation}
T_R(2\pi,\pi;z)=T(0,\pi;z^{-1})\,.
\end{equation}
Demanding then integrability for \eqref{er} we find that \eqref{4} is satisfied for $N(z)=L_{\tau}(0,z)$ and the boundary conditions \cite{Driezen:2019ykp}
\begin{equation}
L_{\tau}(z)|_{\partial\S}=L_{\tau}(z^{-1})|_{\partial\S}\,,
\end{equation}
on both the end points of the string. Using the form of the time component of the Lax pair and expanding order by order in the spectral parameter $z$ we find the following integrable boundary conditions
\begin{equation}\label{13}
\mathcal{B}_+|_{\partial\S}=\mathcal{B}_-|_{\partial\S}\,.
\end{equation}             
\section{Identifying the integrable brane configurations}\label{Ibg}
D-branes are submanifolds which are endowed with a gauge invariant two-form $\omega$. Usually in the literature the starting point to determine a D-brane configuration, i.e. the submanifold and the two-form, is directly from the boundary conditions which it corresponds. This project has been carried out in \cite{Alekseev:1998mc}\cite{Stanciu:1999id}\cite{Stanciu:2000fz}\cite{Figueroa-OFarrill:2000gfl}\cite{Figueroa-OFarrill:1999cmq}, where it was shown that the maximally symmetric boundary conditions
\begin{equation}\label{14}
J_+=\Omega J_-\,, \quad\Omega\in\text{Aut}(G)\,,
\end{equation}
in the WZW model, describe branes which lie in conjugacy classes $\mathcal{C}_{\omega}(f)=\omega(h)fh^{-1}$, with $\omega$ defined as $\omega(e^{tX})=e^{t\Omega[X]}$. In the same line, in \cite{Driezen:2018glg} the authors showed that the integrable boundary conditions 
\begin{equation}\label{15}
A_+=\Omega A_-\,,\quad\Omega\in\text{Aut}(G)\,,
\end{equation}
for the $\l$-deformed sigma-model, describe branes which differ only in terms of size compared to the ones described by \eqref{14}. Furthermore, they derived the boundary two-form for $G=SU(2)$ and $SL(2,\mathbb{R})$, which is $\l$-dependent. Using this form they showed that the points in which the stable D-branes are located are independent of the deformation parameter. 
	
\noindent In our cases the task of determining the brane configurations directly from the integrable conditions \eqref{229}, \eqref{230'}, \eqref{10}, \eqref{13} is more complicated. Thus using a different approach based on \cite{Stanciu:2000fz} \cite{Elitzur:2001qd} \cite{Sark2} \cite{Gawedzki:2001rm}, we will show that the corresponding boundary conditions for the three models under study describe branes whose Dirichlet directions remain invariant when we turn on the deformation parameters. Demanding a well defined action we will determine the induced boundary two-form  for an arbitrary group $G$. In this framework the reason of the stable branes being located at points that are independent of the deformation parameters will become more transparent. Finally, let us mention that the brane geometries we will derive correspond to the cases where the automorphisms appearing in the boundary equations act trivially, i.e. $\Omega=\mathbb{1}$. Furthermore, in Appendix A we show how one can include the case where $\Omega$ acts non trivially.  
\subsection{The isotropic deformation at equal levels}
In the case of a worldsheet with no boundaries the action of model (I) is given in \eqref{16'}. To proceed we recall this action in the form
\begin{equation}\label{19a}
\begin{split}
S_{k;\l_1,\l_2}=\int_{\S}L_{k;\l_1,\l_2}+\int_{M}H_{k;\l_1,\l_2}\,,
\end{split}
\end{equation}
where $M$ is a three manifold bounded by the worldsheet $\S$, i.e. $\partial M=\S$ and \footnote{For the worldsheet conventions see \eqref{lol}.}
\begin{align}\label{526}\nonumber
L_{k,\l_1,\l_2}=-\frac{k}{8\pi}\sum_{i=1}^{2}&\Tr(g_i^{-1}\partial_{\m}g_i,g_i^{-1}\partial^{\m}g_i)+2\l_i\Tr(\partial^{\m}g_ig_i^{-1},\mathcal{O}_{i+1,i}g_{i+1}^{-1}\partial_{\m}g_{i+1})\\
&+2\l_i\l_{i+1}\Tr(\partial^{\m}g_ig_i^{-1},\mathcal{O}_{i+1,i}D_{i+1}^Tg_i^{-1}\partial_{\m}g_i)\,,\\
H_{k,\l_1,\l_2}=\frac{k}{4\pi}\sum_{i=1}^{2}&(H_{WZ}(g_i)+\l_id\Tr(dg_ig_i^{-1}\wedge\mathcal{O}_{i+1,i}(g_{i+1}^{-1}dg_{i+1}+\l_{i+1}D_{i+1}^Tg_i^{-1}dg_i)))\,.\nonumber
\end{align}
As it stands, action \eqref{19a} is not well defined for a worldsheet with boundaries. The ill defined term is the last one as there is no region $M$ bounded by $\S$ when $\S$ itself has boundaries. To fix that, action \eqref{19a} is modified as
\begin{equation}\label{20}
\begin{split}
S_{k;\l_1,\l_2}&=\int_{\S}L_{k;\l_1,\l_2}+\int_{M'}H_{k;\l_1,\l_2}-\int_{D}\omega_{k;\l_1,\l_2}\,,
\end{split}
\end{equation}
where $M'$ is a three-dimensional manifold bounded by $\S\cup D$, with $D$ a two dimensional disc embedded in the brane, and the boundary two-form $\omega_{k;\l_1,\l_2}$ is such that \cite{Alekseev:1998mc}, \cite{Stanciu:2000fz}, \cite{Gawedzki:2001rm}, \cite{Klimcik:1995np}
\begin{equation}\label{21}
H_{k;\l_1,\l_2}|_{\text{brane}}=d\omega_{k;\l_1,\l_2}\,,
\end{equation}
where one restricts to the brane surface. At this point the form of the brane geometry enters into consideration. We will embed in the model a brane whose worldvolume lies in a product of two $G$-conjugacy classes, i.e.
\begin{equation}\label{22}
\mathcal{C}_{f_1f_2}=\mathcal{C}_{f_1}\times\mathcal{C}_{f_2}=\{(h_1f_1h_1^{-1},h_2f_2h_2^{-1}),|\forall h_1,h_2\in G\}\,,
\end{equation}
where $f_1, f_2$ are fixed elements of the group chosen from the Cartan torus of $G$. The dimension of \eqref{22} is $2(d_G-r_G)$, where $d_G=\dim(G)$ and $r_G=\text{rank}(G)$.

\noindent Having specified the brane geometry we are in the position to compute the boundary two form. Using \eqref{21}, \eqref{22} we find that\footnote{Notice that we can add an arbitrary exact two form in \eqref{530} $F=dA$, which in principle can be $\l$-dependent. However, such a choice would not lead to the desired integrable boundary conditions below in \eqref{31}.}
\begin{equation}\label{530}
\begin{split}
\omega_{k;\l_1,\l_2}=\frac{k}{4\pi}&\sum_{i=1}^{2}\Big(\omega_{WZ}(h_i)+\l_i\Tr(dg_ig_i^{-1}\wedge\mathcal{O}_{i+1,i}g_{i+1}^{-1}dg_{i+1})|_{\mathcal{C}_{f_1f_2}}\\
&+\l_i\l_{i+1}\Tr(dg_ig_i^{-1}\wedge \mathcal{O}_{i+1,i}D_{i+1}^Tg_i^{-1}dg_i)|_{\mathcal{C}_{f_1f_2}}\Big)\,,
\end{split}
\end{equation}
where $\omega_{\text{WZ}}(f_i)$ reads \cite{Stanciu:2000fz}\cite{Elitzur:2001qd}\cite{Gawedzki:2001rm}
\begin{equation}\label{24}
\omega_{\text{WZ}}(f_i)=\Tr(h^{-1}_idh_i\wedge f_ih_i^{-1}dh_if_i^{-1}),\quad i=1,2
\end{equation}  
and is such that it satisfies the relation, $H_{WZ}(g_i)|_{\mathcal{C}_{f_1f_2}}=d\omega_{WZ}(h_i)$. The second term in \eqref{530} is just the deformation dependent two form in \eqref{526} restricted on the brane \eqref{22}, i.e. we just replace the group elements $g_i$ with the boundary values, $g_i|_{\partial\S}=h_if_ih_i^{-1}$. Having determined the boundary two-form we proceed by computing the boundary contribution in the variation of \eqref{20}, where we present the technical details in the appendix \ref{bconI}. The result is
 \begin{align}\label{28}\nonumber
 \delta S|_{\partial\S}&=\frac{k}{2\pi}\int_{\partial\S}\Tr(\delta h_1h_1^{-1},\nabla_+g_1g_1^{-1}+g_1^{-1}\nabla_-g_1+(A_{1+}-A_{2+})+(A_{1-}-A_{2-}))\\\nonumber
 &+\frac{k}{2\pi}\int_{\partial\S}\Tr(\delta h_2h_2^{-1},\nabla_+g_2g_2^{-1}+g_2^{-1}\nabla_-g_2+(A_{2+}-A_{1+})+(A_{2-}-A_{1-}))\\
&=\frac{k}{2\pi}\int_{\partial\S}\Tr(\delta h_1h_1^{-1},\l_1^{-1}A_{1+}-\l_2^{-1}A_{2-}-(A_{2+}-A_{1-}))\\\nonumber
&+\frac{k}{2\pi}\int_{\partial\S}\Tr(\delta h_2h_2^{-1},\l_2^{-1}A_{2+}-\l_1^{-1}A_{1-}-(A_{1+}-A_{2-}))\,.
 \end{align}
 To pass from the first to the second equality we used the fact the the fields $A_{i\pm}$ are solutions of the constraint equations
 \begin{equation}\label{B1'}
 \nabla_+g_ig_i^{-1}=(\l_i^{-1}-1)A_{i+},\quad g^{-1}_{i}\nabla_-g_{i}=-(\l_{i+1}^{-1}-1)A_{i+1-},\quad i=1,2\,.
 \end{equation} 
 It is straightforward to see that setting $\l_1=\l_2=\l$, the vanishing of the boundary contribution leads exactly to the integrable conditions in \eqref{229} for trivial automorphisms 
 \begin{equation}\label{31}
 A_{1+}|_{\partial\S}=A_{2-}|_{\partial\S},\quad A_{2+}|_{\partial\S}=A_{1-}|_{\partial\S}\,.
 \end{equation}
 Recall that in section \ref{ide} we derived the above equations using the boundary monodromy method and were led to the same condition $\l_1=\l_2$, demanding that \eqref{227'} satisfy \eqref{222}. 
 Turning our attention to the other set of integrable boundary conditions in \eqref{230'} we embed in \eqref{20} a different brane geometry, known in the literature  as permutation branes, which is defined as
 \begin{equation}\label{32}
 {\mathcal{C}^{\pi}}_{f_1f_2}={\mathcal{C}^{\pi}}_{f_1}\times\left({\mathcal{C}^{\pi}}_{f^{-1}_2}\right)^{-1}=\{(h_1f_1h_2^{-1},h_2f_2h_1^{-1}),| \forall h_1, h_2\in G\}\,.
 \end{equation}
In general the dimension of \eqref{32} is $2d_G-r_G$, but for certain values of $f_1, f_2$ it will have a lower value \cite{Sarkissian:2003yw}. In particular, if $f_1f_2=\mathbb{1}$ then we have $g_1|_{\partial\S}=g_2^{-1}|_{\partial\S}$ on the boundary. In this case the dimension is given by $d_G$. The boundary two form induced in \eqref{32} is given as in \eqref{530}, but with the difference now that\footnote{$\omega_{\text{WZ}}$ is defined such that \cite{Sarkissian:2003yw}
\begin{equation}
(H_{\text{WZ}}(g_1)+H_{\text{WZ}}(g_2))|_{{\mathcal{C}^{\pi}}_{f_1f_2}}=d(\omega_{\text{WZ}}(f_1)+\omega_{\text{WZ}}(f_2)).
\end{equation}} 
 \begin{equation}\label{33}
 \omega_{\text{WZ}}(f_i)=\Tr(h_i^{-1}dh_i\wedge f_ih_{i+1}^{-1}dh_{i+1}f_i^{-1})\,,\quad i=1,2
 \end{equation}  
 and the two-form induced by the deformation is restricted on the brane surface \eqref{32}. As before we are now in position to determine the boundary contribution (see again appendix \ref{bconI}) which reads 
 \begin{equation}\label{35}
 \begin{split}
 \delta S|_{\partial\S}&=\frac{k}{2\pi}\int_{\partial\S}\Tr(\delta h_1h_1^{-1},\nabla_+g_1g_1^{-1}+g_2^{-1}\nabla_-g_2)+\Tr(\delta h_2h_2^{-1},\nabla_+g_2g_2^{-1}+g_1^{-1}\nabla_-g_1)\\
 &=\frac{k}{2\pi}\int_{\partial\S}(\l_1^{-1}-1)\Tr(\delta h_1h_1^{-1},A_{1+}-A_{1-})+(\l_2^{-1}-1)\Tr(\delta h_2h_2^{-1},A_{2+}-A_{2-})\,,
 \end{split}
 \end{equation}
 where we used \eqref{B1'}. Its vanishing leads to the boundary conditions 
 \begin{equation}\label{36}
 A_{1+}|_{\partial\S}=A_{1-}|_{\partial\S}\,,\quad A_{2+}|_{\partial\S}=A_{2-}|_{\partial\S}\,,
 \end{equation}
 which exactly correspond to the other set of integrable conditions in \eqref{229}.
 
 \noindent Thus we found that model (I) admits two kinds of brane configurations  along its RG flow, namely that 
 \begin{equation}\label{B19}
 \begin{split}
 &\mathcal{C}_{f_1f_2}=\{(h_1f_1h_1^{-1},h_2f_2h_2^{-1},|\forall h_1,h_2\in G)\}\,,\\
 \omega_{k;\l_1,\l_2}=&\frac{k}{4\pi}\sum_{i=1}^{2}\Big(\omega_{WZ}(h_i)+\l\Tr(dg_ig_i^{-1}\wedge\mathcal{O}_{i+1,i}g_{i+1}^{-1}dg_{i+1})|_{\mathcal{C}_{f_1f_2}}\\
 &+\l^2\Tr(dg_ig_i^{-1}\wedge \mathcal{O}_{i+1,i}D_{i+1}^Tg_i^{-1}dg_i)|_{\mathcal{C}_{f_1f_2}}\Big)\,,\\
 &\omega_{WZ}(f_i)=\Tr(h^{-1}_idh_i\wedge f_ih_i^{-1}dh_if_i^{-1})\,,
 \end{split}
 \end{equation}
 and that
\begin{equation}\label{B20}
\begin{split}
&{{\mathcal{C}^{\pi}}}_{f_1f_2}=\{(h_1f_1h_2^{-1},h_2f_2h_1^{-1},|\forall h_1,h_2\in G)\}\,,\\
\omega_{k;\l_1,\l_2}=&\frac{k}{4\pi}\sum_{i=1}^{2}\Big(\omega_{WZ}(h_i)+\l_i\Tr(dg_ig_i^{-1}\wedge\mathcal{O}_{i+1,i}g_{i+1}^{-1}dg_{i+1})|_{\mathcal{C}_{\pi}}\\
&+\l_i\l_{i+1}\Tr(dg_ig_i^{-1}\wedge \mathcal{O}_{i+1,i}D_{i+1}^Tg_i^{-1}dg_i)|_{\mathcal{C}_{\pi}}\Big)\,,\\
&\omega_{WZ}(f_i)=\Tr(h^{-1}_idh_i\wedge f_ih_{i+1}^{-1}dh_{i+1}f_i^{-1})\,,
\end{split}
\end{equation}
As we have said \eqref{B19} require $\l_1=\l_2=\l$, while \eqref{B20} do not. Moreover, they are solutions of the boundary conditions \eqref{31} and \eqref{36} respectively, thus proving that they preserve integrability of the model for arbitrary values of the couplings.
 \subsection{The isotropic deformation at unequal levels}
 The action of model (II) \eqref{30a'}, which describes the deformation of the direct product of two current algebras, in the precence of boundaries can be written as
 \begin{equation}\label{320}
 \begin{split}
 S_{k_1,k_2;\l_1,\l_2}&=\int_{\S}L_{k_1,k_2;\l_1,\l_2}+\int_{M'}H_{k_1,k_2;\l_1,\l_2}-\int_{D}\omega_{k_1,k_2;\l_1,\l_2}\,,
 \end{split}
 \end{equation}
 where 
\begin{align}\label{542}\nonumber
L_{k_1,k_2,\l_1,\l_2}=-\frac{1}{8\pi}\sum_{i=1}^{2}&k_i\Tr(g_i^{-1}\partial_{\m}g_i,g_i^{-1}\partial^{\m}g_i)+2\l_ik^{(i+1)}\Tr(\partial^{\m}g_ig_i^{-1},\mathcal{O}_{i+1,i}g_{i+1}^{-1}\partial_{\m}g_{i+1})\\
&+2\l_i\l_{i+1}k_i\Tr(\partial^{\m}g_ig_i^{-1},\mathcal{O}_{i+1,i}D_{i+1}^Tg_i^{-1}\partial_{\m}g_i))\\
H_{k_1,k_2,\l_1,\l_2}=\frac{1}{4\pi}\sum_{i=1}^{2}(k_i&H_{WZ}(g_i)+\l_id\Tr(dg_ig_i^{-1}\wedge\mathcal{O}_{i+1,i}(k^{(i+1)}g_{i+1}^{-1}dg_{i+1}+k_i\l_{i+1}D_{i+1}^Tg_i^{-1}dg_i)))\,.\nonumber
\end{align}
with $k^{(i)}=\sqrt{k_ik_{i+1}}$.
The integrability preserving boundary conditions consistent with the model are of the form \eqref{10}, thus the brane we will consider is \eqref{22}. As before the boundary two-form trivializing the three form $H_{k_1,k_2;\l_1,\l_2}$ on the brane $\mathcal{C}_{f_1f_2}$ is
\begin{equation}\label{543}
\begin{split}
\omega_{k_1,k_2;\l_1,\l_2}=\frac{1}{4\pi}\sum_{i=1}^{2}&\Big(k_i\omega_{WZ}(h_i)+\l_ik^{(i+1)}\Tr(dg_ig_i^{-1}\wedge\mathcal{O}_{i+1,i}g_{i+1}^{-1}dg_{i+1})|_{\mathcal{C}_{f_1f_2}}\\
&+\l_i\l_{i+1}k_i\Tr(dg_ig_i^{-1}\wedge \mathcal{O}^T_{i+1,i}D_{i+1}^Tg_i^{-1}dg_i)|_{\mathcal{C}_{f_1f_2}}\Big)\,.
\end{split}
\end{equation}
where $\omega(f_i)$ is given in \eqref{24}. Since the technical details are the same with the previous case we will just write down the result  
 \begin{align}\label{B27}\nonumber
 \delta S|_{\partial\S}&=\frac{k_1}{2\pi}\int_{\partial\S}\Tr(\delta h_1h_1^{-1},\nabla_+g_1g_1^{-1}+g_1^{-1}\nabla_-g_1+(A_{1+}-A_{2+})+(A_{1-}-A_{2-}))\\\nonumber
 &+\frac{k_2}{2\pi}\int_{\partial\S}\Tr(\delta h_2h_2^{-1},\nabla_+g_2g_2^{-1}+g_2^{-1}\nabla_-g_2+(A_{2+}-A_{1+})+(A_{2-}-A_{1-}))\\
 &=\frac{k_1}{2\pi}\int_{\partial\S}\Tr(\delta h_1h_1^{-1},\l_0^{-1}(\l_1^{-1}A_{1+}-\l_2^{-1}A_{2-})-(A_{2+}-A_{1-}))\\\nonumber
 &+\frac{k_2}{2\pi}\int_{\partial\S}\Tr(\delta h_2h_2^{-1},\l_0(\l_2^{-1}A_{2+}-\l_1^{-1}A_{1-})-(A_{1+}-A_{2-}))\,.
 \end{align}
As before to pass from the first to second equality we used the constraints
\begin{equation}\label{B22}
\begin{split}
&\nabla_+g_1g_1^{-1}=(\l_0^{-1}\l_1^{-1}-1)A_{1+}\,,\quad g^{-1}_2\nabla_-g_2=-(\l_0\l_1^{-1}-1)A_{1-}\,,\\
&\nabla_+g_2g_2^{-1}=(\l_0\l_2^{-1}-1)A_{2+}\,,\quad g^{-1}_1\nabla_-g_1=-(\l_0^{-1}\l_2^{-1}-1)A_{2-}\,.
\end{split}
\end{equation}
Setting $\l_1=\l_2=\l$ one can see that the vanishing  of \eqref{B27} leads to the desired integrable boundary conditions found in \eqref{10} 
 \begin{equation}\label{B28}
 A_{1+}|_{\partial\S}=A_{2-}|_{\partial\S}\,,\quad A_{2+}|_{\partial\S}=A_{1-}|_{\partial\S}\,.
 \end{equation}
Summarizing our results, the model (II) admits the following integrable brane configuration along its RG flow
 \begin{equation}\label{B19'}
\begin{split}
&\mathcal{C}_{f_1f_2}=\{(h_1f_1h_1^{-1},h_2f_2h_2^{-1},|\forall h_1,h_2\in G)\}\,,\\
\omega_{k_1,k_2;\l_1,\l_2}=\frac{1}{4\pi}&\sum_{i=1}^{2}\Big(k_i\omega_{WZ}(h_i)+\l k^{(i+1)}\Tr(dg_ig_i^{-1}\wedge\mathcal{O}_{i,i+1}g_{i+1}^{-1}dg_{i+1})|_{\mathcal{C}_{f_1f_2}}\\
&+\l^2k_i\Tr(dg_ig_i^{-1}\wedge \mathcal{O}_{i,i+1}D_{i+1}^Tg_i^{-1}dg_i)|_{\mathcal{C}_{f_1f_2}}\Big)\,,\\
&\omega_{WZ}(f_i)=\Tr(h^{-1}_idh_i\wedge f_ih_i^{-1}dh_if_i^{-1})\,.
\end{split}
\end{equation}  
We remind to the reader that the model (II) interpolates between two exact CFTs in the UV and IR fixed points for $\l=0$ and $\l=\l_0$ respectively, and that the integrable branes $\eqref{B19'}$ interpolate between two conformal configurations at the same points, preserving one copy of the Virassoro algbera respectively.\footnote{Recall that the boundary conditions \eqref{B28} satisfy \eqref{222} for arbitrary values of the couplings. Thus at the conformal points the two copies of the Virassoro algebra are reduced to one.}

\noindent Turning our attention to the other set of integrable boundary conditions \eqref{237} we notice that they suggest that the corresponding brane geometry contains a permutation between the two groups like in \eqref{32}.\footnote{To see this note that in the UV limit the boundary conditions \eqref{32} become
\begin{equation}
k_1J_{1+}|_{\partial\S}=-k_2J_{2-}|_{\partial\S}\,,\quad k_2J_{2+}|_{\partial\S}=-k_1J_{1-}|_{\partial\S}\,.
\end{equation}} However, permutation branes can not be defined consistently in the case of $k_1\neq k_2$, in the sense that there is no two-form trivializing the three-form $H_{k_1,k_2;\l_1,\l_2}$ on the brane, leading to the conclusion that \eqref{237} do not admit a geometric solution.
 \subsection{The isotropic deformation of $G_{k_1}\times G_{k_2}/G_{k_1+k_2}$ space}
 As a final example we consider the $\l$-deformed of a $G_{k_1}\times G_{k_2}/G_{k_1+k_2}$ coset space \eqref{231'}.
 Repeating the same steps as before we rewrite its action as
  \begin{equation}\label{328}
 S_{k_1,k_2;\l}=\int_{\S}L_{k_1,k_2;\l}+\int_{M}H_{k_1,k_2;\l}-\int_{D}\omega_{k_1,k_2;\l}\,,
 \end{equation}
 with
 \begin{align}\label{576'}\nonumber
 &L_{k_1,k_2,\l}=-\frac{1}{8\pi}\sum_{i=1}^{2}k_i\Tr(g_i^{-1}\partial_{\m}g_i,g_i^{-1}\partial^{\m}g_i)-2\l s_is_{i+1}k_i\Tr(\partial^{\m}g_ig_i^{-1},\L_{i,i+1}^{-T}(D_{i+1}^T-\mathbb{1})g_i^{-1}\partial_{\m}g_i)\\\nonumber
 &\hspace{18mm}+2(1-\l)k_i\Tr(\partial^{\m}g_ig_i^{-1},\L^{-T}_{i,i+1}(s_ig^{-1}_i\partial_{\m}g_i+s_{i+1}g^{-1}_{i+1}\partial_{\m}g_{i+1}))\,,\\\nonumber
 &H_{k_1,k_2,\l}=\frac{1}{4\pi}\sum_{i=1}^{2}(k_iH_{WZ}(g_i)-2\l s_is_{i+1}k_id\Tr(dg_ig_i^{-1},\L_{i,i+1}^{-T}(D_{i+1}^T-\mathbb{1})g_i^{-1}dg_i))\\
 &\hspace{18mm}+2(1-\l)k_id\Tr(dg_ig_i^{-1}\wedge\L^{-T}_{i,i+1}(s_ig^{-1}_idg_i+s_{i+1}g^{-1}_{i+1}dg_{i+1}))\,.
 \end{align}
If we consider the brane \eqref{22} and the boundary two-form trivializing $H_{k_1,k_2;\l}$ on the brane, we find that the boundary contribution to the variation of \eqref{328}  is
 \begin{align}\label{230}\nonumber
 \delta S|_{\partial\S}&=\frac{1}{2\pi}\int_{\partial\S}k_1\Tr(\delta h_1h_1^{-1},\nabla_+g_1g_1^{-1}+g^{-1}_1\nabla_-g_1)+k_2\Tr(\delta h_2h_2^{-1},\nabla_+g_2g_2^{-1}+g^{-1}_2\nabla_-g_2)\\
 &=\frac{k(\l^{-1}-1)}{4\pi}\int_{\partial\S}\Tr(\delta h_1h_1^{-1},\mathcal{B}_+-\mathcal{B}_-)-\Tr(\delta h_2h_2^{-1},\mathcal{B}_+-\mathcal{B}_-)\,,
 \end{align}
where recall that $k=k_1+k_2$ and $s_i=k_i/k$, $i=1,2$. We used also the constraints
\begin{equation}\label{c8}
\begin{split}
&s_1\nabla_+g_1g_1^{-1}=\frac{1}{2}(\l^{-1}-1)\mathcal{B}_+\,,\quad s_2\nabla_+g_2g_2^{-1}=-\frac{1}{2}(\l^{-1}-1)\mathcal{B}_+\,,\\ 
&s_1g_1^{-1}\nabla_-g_1=-\frac{1}{2}(\l^{-1}-1)\mathcal{B}_-\,,\quad s_2g_2^{-1}\nabla_-g_2=\frac{1}{2}(\l^{-1}-1)\mathcal{B}_-\, .
\end{split}
\end{equation} 
The vanishing of \eqref{230} leads to the integrable boundary conditions \eqref{13}. Thus we find that the brane configuration
 \begin{align}\label{B19''}\nonumber
 &\mathcal{C}_{f_1f_2}=\{(h_1f_1h_1^{-1},h_2f_2h_2^{-1},|\forall h_1,h_2\in G)\}\,,\\\nonumber
 \omega_{k_1,k_2;\l}=\frac{1}{4\pi}\sum_{i=1}^{2}&\Big(k_i\omega_{WZ}(h_i)+2k_i(1-\l)\Tr(dg_ig_i^{-1},\L^{-T}_{i,i+1}(s_ig_{i}^{-1}dg_{i}+s_{i+1}g_{i+1}^{-1}dg_{i+1}))|_{\mathcal{C}_{f_1f_2}}\\
 &-2\l s_is_{i+1}k_i\Tr(dg_ig_i^{-1},\L_{i,i+1}^{-T}(D_{i+1}^T-\mathbb{1})g_i^{-1}dg_i)|_{\mathcal{C}_{f_1f_2}})\,,\\
 &\omega_{WZ}(f_i)=\Tr(h^{-1}_idh_i\wedge f_ih_i^{-1}dh_if_i^{-1})\,,\nonumber
 \end{align}
 preserves integrability of the deformed coset space. In the following sections we will consider a different brane geometry, known in the literature as generalized permutation branes \cite{Fredenhagen:2005an}, and we will show that it also solves the boundary conditions \eqref{13}.
 \subsection{The allowed set of integrable branes}\label{Qc}
 As discussed in the literature \cite{Alekseev:1998mc}\cite{Stanciu:2000fz}\cite{Elitzur:2001qd}\cite{Gawedzki:1999bq}\cite{Gawedzki:2001rm}, in the Lagrangian description of boundary sigma-models the absence of topological obstructions, originating from WZ-type terms, requires that the relative cohomology class $[(H,\omega)]$ should be integral, that is  
 \begin{equation}\label{q}
\int_{S^3}H-\int_{S^2}\omega\in 2\pi\mathbb{Z}\,,
 \end{equation}
where $S^3$ is a three dimensional ball, whose boundary $S^2$ is mapped into the brane. Picking the specific form of the three-form $H$ for each model, given in \eqref{526}, \eqref{542}, \eqref{576'} and the corresponding boundary two-forms $\omega$ in \eqref{B19}, \eqref{B20}, \eqref{B19'}, \eqref{B19''}, we find that \eqref{q} is independent of the deformation parameters and specifically equals to 
  \begin{equation}\label{qq}
 \sum_{i=1}^{2}\frac{k_i}{4\pi}\left(\int_{S^3}H_{\text{WZ}}(g_i)-\int_{S^2}\omega_{\text{WZ}}(f_i)\right)\in 2\pi\mathbb{Z}\,.
 \end{equation}
 In other words, the flux of the $F$ field, $F=dA=\omega-B$, through any two sphere $S^2$ is quantized and independent of the deformation. 
Equation \eqref{qq} leads to the quantization of the elements $f$ chosen from the Cartan torus of the group $G$. This project has been carried out for the branes under study for compact simply connected groups, see e.g. \cite{Elitzur:2001qd,Gawedzki:1999bq,Stanciu:2001vw}.

\noindent To be more specific let us consider the case where $G=SU(2)$, and we choose the convenient parametrization \begin{equation}\label{A2}
g(\psi,\theta,\phi)=
\begin{pmatrix}
&\cos(\psi)+i\cos(\theta)\sin(\psi)&\sin(\psi)\sin(\theta)e^{i\phi}\\
&-\sin(\psi)\sin(\theta)e^{-i\phi}&\cos(\psi)-i\cos(\theta)\sin(\psi)
\end{pmatrix}\,,
\end{equation}
with $\theta\in[0,\pi]$, $\phi\in[0,2\pi]$ and $\psi\in[0,\pi]$.

\noindent Due to the previous analysis, the model (I) admits two kinds of stable integrable branes along its RG flow given in \eqref{B19} and \eqref{B20}. In the parametrization \eqref{A2} and using \eqref{qq} we find that the first set of branes are $(k+1)^2$ hypersurfaces located at \cite{Alekseev:1998mc}, \cite{Stanciu:2000fz}, \cite{Bachas:2000ik}, \cite{Gawedzki:1999bq},  
\begin{equation}\label{235'}
\psi_i=\frac{n_i\pi}{k}\,,\quad n_i=0,1\,,\dots,k,\quad i=1,2\,,
\end{equation}
Turning our attention to the other set of integrable branes, \eqref{B20}, we know from the analysis in \cite{Sarkissian:2003yw} that the number of stable permutation branes in a product of WZW models, $G_k\times G_k$, equals the number of maximally symmetric branes in one copy $G_k$. Since this picture persists along the whole deformation line, model (I) admits additionally $(k+1)$ stable permutation branes embedded in its $\l$-dependent target space.

\noindent Similarly, model (II) admits $(k_1+1)(k_2+1)$ integrable branes described by \eqref{B19'}, which compared to \eqref{235'} are located at
\begin{equation}\label{235}
\psi_i=\frac{n_i\pi}{k_i}\,,\quad n_i=0,1,\dots,k_i\,,\quad i=1,2\,.
\end{equation}
The same result \eqref{235} holds for the integrable branes of model (III) given in \eqref{B19''}.

\noindent It is important to notice, that even though the bulk degrees of freedom of a field theory reduce in number along its RG flow, from the UV towards the IR CFT, the allowed branes remain invariant. 
\subsection{Comments on the brane geometry}
The geometry of the target spaces of the models under study is $\l$-dependent. However, we saw that the integrable branes defined in each model, have some interesting $\l$-independent features. They are hypersurfaces located at points in the transverse space which remain fixed along the corresponding RG flows, i.e. their Dirichlet directions are independent of the deformation. This feature were also found in \cite{Driezen:2018glg}, where additionally the authors showed that the deformation affects only the size of the integrable branes. Here we will give a plausible argument that this simple action of the deformation does not hold for the highest dimensional integrable branes that the corresponding target spaces admit.   

\noindent We will choose to demonstrate our arguments for the branes \eqref{22} embedded in model (II) for $G=SU(2)$. In the UV they are presented in table \eqref{table:eeeeee'}.\footnote{The conjugacy classes of $SU(2)$ are two points at $\psi=0,\pi$ and $k-1$ $S^2$-spheres located at $\psi=n\pi/k$ for $n=1,\dots, k-1$.} 
 \begin{table}[h!]
 	\begin{footnotesize}
 		\renewcommand{\arraystretch}{2.7}
 		\begin{center}
 			\begin{tabular}{ |p{3.3cm}|p{3.5cm}|p{2.6cm}| }
 				\hline
 				Number $(N)$ & Geometry & Dimension\\
 				\hline
 				$N_1=4$   &$\{\pm\mathbb{1}\}\times\{\pm\mathbb{1}\}$  &$0$\\
 				\hline
 				$N_2=2(k_1+k_2-2)$ &$S^2\times\{\pm\mathbb{1}\}$, $\{\pm\mathbb{1}\}\times S^2$ & $2$\\
 				\hline
 				$N_3=(k_1-1)(k_2-1)$ &$S^2\times S^2$ &$4$\\
 				\hline
 			\end{tabular}
 		\end{center}
 		\caption{\footnotesize{
 		The product of conjugacy classes in $SU(2)_{k_1}\times SU(2)_{k_2}$. Note that $N_1+N_2+N_3=(k_1+1)(k_2+1)$.}}
 		\label{table:eeeeee'}
 	\end{footnotesize}
 \end{table}  
Turning on the deformation, the topology of the $D2$-branes is unaffected. Specifically the $D2$-branes change only in terms of size as in \cite{Driezen:2018glg}. This can be seen by restricting the background metric to $\psi_i=n_i\pi/k_i$ with  $n_i=1,\dots,k_i-1$ and $\psi_{i+1}=0,\pi$,
\begin{equation}
d\hat{s}^2_i=\frac{k_i(1-\l^4)\sin^2\psi_i(d\theta_i^2+\sin^2\theta_id\phi_i^2)}{1+\l^4-2\l^2\cos2\psi_i}\,,\quad g_{i+1}=\pm\mathbb{1}\,,\quad i=1,2\,,
\end{equation}
What about the rest $N_3=(k_1-1)(k_2-1)$ integrable branes? Setting $\psi_i=n_i\pi/k_i$ with $n_i\neq 0,\pi$ for both $i=1,2$, we find that they are four dimensional hypersurfaces, $\mathcal{S}_{\psi_1\psi_2}=\mathcal{S}_{n_1n_2}(\theta_1,\phi_1,\theta_2,\phi_2)$, with the coordinates parameterizing them coupled among themselves in a non trivial $\l$-dependent way. Due to the complexity of the induced metric we do not present it here. However, calculating the Gauss-Bonnet term we found that the topology of the corresponding branes is invariant and equals that of a product of two-spheres.
\subsection{Some comments on integrable brane geometries for $N>2$}
Let us consider the $\l$-deformation of $N$ WZW models all at the same level $k$ \cite{Georgiou:2017oly}. In that work it was shown that it admits $N$ independent Lax pairs of the form \eqref{27} where the fields $A_{i\pm}$ are given in terms of $g_i$, $i=1,\dots, N$ as
\begin{equation}
\begin{split}
&A_{i+}=(\mathbb{1}-\hat{D}_i\dots\hat{D}_{N+i-1})^{-1}\sum_{j=i}^{N+i-1}\hat{D}_j\dots\hat{D}_{j-1}\l_jJ_{j+}\,,\\
&A_{i-}=-(\mathbb{1}-\hat{D}^T_{i-1}\dots\hat{D}_{N+i-2})^{-1}\sum_{j=i-1}^{N+i-2}\hat{D}^T_j\dots\hat{D}^T_{j-1}\l_jJ_{j-}\,,
\end{split}
\end{equation}
where $\hat{D}_i=\l_iD_i$. The boundary conditions preserving its integrable structure can be put in the form
\begin{equation}
\mathbb{A}_+|_{\partial\S}=\mathbb{R}_N\mathbb{A}_-|_{\partial\S}\,,
\end{equation}
where $\mathbb{A}_{\pm}^T=(A_{1\pm},A_{2\pm},\dots, A_{N\pm})$ and $\mathbb{R}_N$ is the gluing matrix describing all the possible conditions between the $A_{i\pm}$'s on the boundary.

\noindent The simplest case for $N=3$ suffices to deduce the general structure. Following the same procedure as in section \ref{Ibc} one can show that all possible integrable conditions are described by gluing matrices of the form
 \begin{equation}\label{bc'}
 \mathbb{R}_3=\begin{pmatrix}
 \Omega_1&0&0\\
 0&\Omega_2&0\\
 0&0&\Omega_3
 \end{pmatrix}\,,
 \end{equation}
by which only the components of the same gauge field are related on the boundary, and of the form 
\begin{equation}\label{bc}
 \mathbb{R}_3=\begin{pmatrix}
\Omega_1&0&0\\
0&0&\Omega\\
0&\Omega^{-1}&0
\end{pmatrix}\,,
\end{equation}
by which components of different gauge fields are related as well. Two more gluing matrices similar to \eqref{bc} are also allowed by performing a cyclic permutation of the indices $(1,2,3)$. Moreover, due to consistency reasons explained in section \ref{Ibc} and the demand of a vanishing momentum flow across the boundary we find that $\mathbb{R}_3^2=\mathbb{1}$ and $\mathbb{R}_3^T=\mathbb{R}_3^{-1}$. For simplicity, we will consider trivial acting automorphisms, i.e. $\Omega_i=\mathbb{1}$.

\noindent Since we want to study cases with arbitrary $N$ we found it convenient to relate the matrices $\mathbb{R}_N$ to the elements of the symmetry group $S_N$. In the $N=3$ case, labeling
every set $A_{i\pm}$ with $i=1,2,3$ we find that $\mathbb{R}_3$ corresponds to elements of $S_3$ of the form $(i)(j)(k)$, i.e. a product of three distinct one-cycles, and $(i)(jk)$, i.e. a product of a one-cycle and a two-cycle. Specifically the rule is: 
\begin{equation*}
\begin{split}
&(i)\leftrightarrow A_{i+}|_{\partial\S}=A_{i-}|_{\partial\S}\,,\\
&(ij)\leftrightarrow A_{i+}|_{\partial\S}=A_{j-}|_{\partial\S}\,,\quad A_{j+}|_{\partial\S}=A_{i-}|_{\partial\S}\,.
\end{split}
\end{equation*}
The integrable brane geometries corresponding to the boundary conditions \eqref{bc'} and \eqref{bc} are\footnote{The permutation branes between $N$ group elements $g_i$ are defined as
	\begin{equation}
	{\mathcal{C}^{\pi}}_{f_1,\dots,f_i,\dots f_N}=(h_1f_1h_2^{-1},\dots ,h_if_ih_{i+1}^{-1},\dots,h_Nf_Nf_1^{-1})
	\end{equation}}
\begin{equation}\label{343}
\begin{split}
&(1)(2)(3):\quad{\mathcal{C}^{\pi}}_{f_1\,f_2\,f_3}\,,\\
&(1)(23):\quad\left(\mathcal{C}_{f_2}\,,\,{\mathcal{C}^{\pi}}_{f_1\,f_3}\right)\,.
\end{split}
\end{equation}
Thus we see that the boundary conditions consisting of three one-cycles correspond to permutation branes defined on all three group elements, while the other boundary conditions to a conjugacy class times a permutation brane. We would like to see if this correspondence between elements of the symmetric group and the integrable brane geometry holds for $N>3$. Towards this it is straightforward to see that different sets of boundary conditions are classified as $\{n,m\}$ where $n$ is the number of two-cycles and $m$ the number of one-cycles with $N=2n+m$. From combinatorics each class has $N!/(n!m!2^n)$ elements.

\noindent Thus for $N=4$ we have the classes, $\{0,4\}, \{1,2\}, \{2,0\}$. In this case we find that each class does not admit the same brane geometry. Specifically we find that:
\begin{equation}
\begin{split}
&(0,4):\quad{\mathcal{C}^{\pi}}_{f\,f\,f\,f}\,,\\
&(1,2):\quad\left(\mathcal{C}_f,{\mathcal{C}^{\pi}}_{f\,f\,f}\right), \left({\mathcal{C}^{\pi}}_{f\,f},{\mathcal{C}^{\pi}}_{f\,f}\right),\\
&(2,0):\quad\left(\mathcal{C}_f,\mathcal{C}_f,{\mathcal{C}^{\pi}}_{f\,f}\right),{\mathcal{C}^{\pi}}_{f\,f\,f\,f}\,.
\end{split}
\end{equation}
Hence, to generate a general rule we need more involved combinatorics in each class, which we have not performed. However let us mention that every two-cycle between adjacent indices, i.e. $(i,i+1)$, results to a conjugacy class, thus the maximum number of conjugacy classes contained in an integrable brane configuration is $N/2$ for $N$ even and $(N-1)/2$ for $N$ odd.  
\section{The integrable branes in the CFT limits}\label{cftlimits}
\subsection{Integrable branes in the conformal limits of model (II)}\label{cft}
As has been analyzed in \cite{Georgiou:2017jfi}, \cite{Georgiou:2020eoo} model (II) flows from a UV point for $(\l_1,\l_2)=(0,0)$ towards a far IR for $(\l_1,\l_2)=(\l_0,\l_0)$. The conformal symmetry of the UV CFT is based on a product of two current algebra symmetries 
\begin{equation}\label{311}
(G_{k_1}\times G_{k_2})_L\otimes(G_{k_1}\times G_{k_2})_R\,,
\end{equation}
 while the IR CFT is a product of coset and current algebra symmetries, i.e. 
\begin{equation}\label{312}
\left(\frac{G_{k_1}\times G_{k_2-k_1}}{G_{k_2}}\times G_{k_2-k_1}\right)_L\otimes\left(\frac{G_{k_1}\times G_{k_2-k_1}}{G_{k_2}}\times G_{k_2-k_1}\right)_R.
\end{equation}
Keeping the discussion general, conformal boundary conditions reduce the initial symmetry group of a CFT to a large enough subgroup which guarantees conformal invariance of the model. Here we will find the symmetries of model (II) at its UV and IR fixed points under the presence of the branes \eqref{22}. Specifically we will determine the subgroups of the initial symmetry groups  \eqref{311} and \eqref{312} preserved by the corresponding branes.

\noindent The integrable branes \eqref{22}, in the UV limit of model (II), preserve on the boundary the diagonal subgroups of each copy of the chiral symmetry \eqref{311} \cite{Elitzur:2001qd}. Put it plainly the symmetry
\begin{equation}\label{313}
(g_1,g_2)\mapsto (k_{1L}(\s_-)g_1k_{1R}(\s_{+}),k_{2L}(\s_-)g_2k_{2R}(\s_+))\,,
\end{equation}
of two WZW models, is reduced to 
\begin{equation} \label{313'}
(g_1,g_2)|_{\partial\S}\mapsto (k_1(\tau)g_1k^{-1}_1(\tau),k_2(\tau)g_2k^{-1}_2(\tau))\,,
\end{equation}
in the presence of \eqref{22}.   
This symmetry can be realized also by noticing that in the UV limit the boundary conditions \eqref{31} identify the generators of the chiral transformation \eqref{313} of each of the group, i.e.
\begin{equation}\label{314}
J_{1+}|_{\partial\S}=-J_{1-}|_{\partial\S}\,,\quad J_{2+}|_{\partial\S}=-J_{2-}|_{\partial\S}\,.
\end{equation} 
Turning our attention to the IR limit of the flow we see that the integrable conditions \eqref{31} become 
\begin{equation}\label{315}
A_{1+}(\l_0)|_{\partial\S}=A_{2-}(\l_0)|_{\partial\S}\,,\quad A_{2+}(\l_0)|_{\partial\S}=A_{1-}(\l_0)|_{\partial\S}\,,
\end{equation}
where we remind to the reader that the presence of the branes \eqref{22} restrict the deformation parameters to be equal, $\l_1=\l_2=\l$ and we emphasize the fact the fields $A_{1\pm}, A_{2\pm}$ are evaluated on the fixed point $\l=\l_0$. According to \cite{Georgiou:2020eoo} the fields $A_{2+}, A_{1-}$ satisfy a $\hat{\mathfrak{g}}_{k_2-k_1}\oplus\hat{\mathfrak{g}}_{k_2-k_1}$ algebra corresponding to the chiral symmetry
\begin{equation}\label{316}
(G_{k_2-k_1})|_{L}\otimes(G_{k_2-k_1})|_{R}\,, 
\end{equation} 
of the IR CFT. From \eqref{315} the left moving generators $A_{1-}$ are identified with the right moving $A_{2+}$, thus proving that the branes \eqref{22} preserve on the boundary the diagonal subgroup of \eqref{316}. What about the other set of conformal boundary conditions in \eqref{315}? As has been also noted in \cite{Georgiou:2020eoo} the action \eqref{30a'} in the IR limit and under the group redefinition $(g_1,g_2)\to (g'_1=g_1g_2,g'_2=g_2)$ is identified as the gauge fixed version, $g'_3=\mathbb{1}$, of an action which realizes the conformal symmetry \eqref{312} generated by the transformations
\begin{equation}\label{38}
(g'_1,g'_2,g'_3)\mapsto (Lg'_1L^{-1},g'_2L^{-1},Lg'_3),\quad L=L(\s_+,\s_-)\,,
\end{equation}
and
\begin{equation}\label{39}
(g'_2,g'_3)\mapsto (k_L(\s_+)g'_2,g'_3k_R(\s_-))\,.
\end{equation} 
The branes \eqref{22}, which under the group redefinition become $(\mathcal{C}_{f_1}\mathcal{C}_{f_2},\mathcal{C}_{f_2})$, can be viewed as the gauge fixed version of a brane geometry which when embedded in the extended action preserves the gauge symmetry \eqref{38} and the diagonal subgroup of \eqref{39}. Thus the boundary conditions \eqref{315} are the gauge fixed version of the conditions which correspond to the preservation of $\eqref{38}$ and \eqref{39}. We were not able to find the gauge invariant extension of the $(\mathcal{C}_{f_1}\mathcal{C}_{f_2},\mathcal{C}_{f_2})$ geometry.
\subsection{Integrable branes in the conformal limits of model (III) }
As we have mentioned model (III) is invariant under the adjoint action  of $G$ \eqref{118}. It flows from a UV CFT for $\l=0$ towards a CFT in the IR for $\l=1/(s_2-3s_1)$ \cite{Sfetsos:2017sep}. That is
\begin{equation}\label{311'}
\frac{G_{k_1}\times G_{k_2}}{G_{k_1+k_2}}\Longrightarrow\frac{G_{k_1-k_2}\times G_{k_2}}{G_{k_1}}\,.
\end{equation}
The presence of the branes \eqref{B19''} along \eqref{311'} does not spoil gauge invariance of the model. To see this notice that under \eqref{118} the action \eqref{328} transforms 
\begin{equation}\label{557}
\delta_{V}S=\delta_V S|_{\S}+\delta_VS|_{\partial\S}\,,
\end{equation}
where $\delta_V$ stands for the infinitesimal version of \eqref{118}, with the boundary term given by
\begin{equation}\label{557'}
\delta_VS|_{\partial\S}=\frac{k(\l^{-1}-1)}{4\pi}\int_{\partial\S}\Tr(\delta_V h_1h_1^{-1},\mathcal{B}_+-\mathcal{B}_-)-\Tr(\delta_V h_2h_2^{-1},(\mathcal{B}_+-\mathcal{B}_-))\,.
\end{equation}
To find explicitly the contribution \eqref{557'} notice that the geometry of the brane remains invariant, in the sense that
\begin{equation}\label{558}
(g_1,g_2)|_{\partial\S}\mapsto (L^{-1}g_1L,L^{-1}g_2L)|_{\partial\S}=(h'_1f_1{h_1'}^{-1},h'_2f_2{h_2'}^{-1})\,,
\end{equation}
with $h_i'=L^{-1}h_i$. Thus setting $\delta_Vh_1h_1^{-1}=\delta_Vh_2h_2^{-1}$ in \eqref{557'} we find that $\delta_VS|_{\partial\S}=0$, while of course $\delta_VS|_{\S}=0$. This proves that the boundary action \eqref{328} remains invariant. Thus, projecting the $2(d_G-r_G)$-dimensional branes in the deformed coset space we end up with $(d_G-2r_G)$-dimensional integrable branes. In the case $G=SU(2)$ our branes are either $0$ or $1$-dimensional.

\noindent Let us now turn our attention to the conformal limits of \eqref{311'} for $SU(2)$. In \cite{Wurtz:2005mt} the authors have studied the $D$-branes in a diagonal coset space CFT of the form $SU(2)_{k}\times SU(2)_{l}/SU(2)_{m}$, with $m=k+l$. They have argued that the conformal branes are given as a projected product of the $SU(2)$-conjugacy classes on the coset space, i.e.
\begin{equation}\label{381}
\p_{\text{Ad}}(\mathcal{C}_{a_1a_2a_3})=\p_{\text{Ad}}((\mathcal{C}_{a_1}\mathcal{C}_{a_3},\mathcal{C}_{a_2}\mathcal{C}_{a_3}))\,,
\end{equation} 
 where $\pi_{Ad}$ denotes the projector and the numbers $a_1,a_2,a_3$ parametrize the elements of the Cartan torus of $SU(2)$, i.e. $f_i=f_i(a_i)\in SU(2)$, and correspond to the levels $k,l,m$, respectively. Furthermore, they classified the dimensions of the branes \eqref{381} in terms of the values of the $a_i's$ and found them to be $0,1$ and $3$-dimensional.
 
 \noindent Returning to our integrable branes \eqref{B19''} for $G=SU(2)$, corresponding to $D0$- and $D1$-branes, it is straightforward to see that in the UV limit of the flow \eqref{311'} they belong in the class \eqref{381} for
 \begin{equation}\label{kla}
 a_1=\psi_1=\frac{n_1\pi}{k_1}\,,\quad a_2=\psi_2=\frac{n_2\pi}{k_2}\,,\quad a_3=0\,,
 \end{equation}
 with $n_i=0\,,\dots,k_i\,,\,i=1,2$.  
 The IR CFT in \eqref{311'} is realized under the coordinate transformation $(g_1,g_2)\to (g'_1=g_1,g'_2=g_2g_1)$ \cite{Sfetsos:2017sep}. In these coordinates our integrable branes are written as 
 \begin{equation}\label{561}
 ({g'}_1,{g'}_2)|_{\partial\S}=(\mathcal{C}_{\psi_1},\mathcal{C}_{\psi_2}\mathcal{C}_{\psi_1})\,.
 \end{equation}
 Thus  in the IR they belong again in the class \eqref{381} for
 \begin{equation}
 a_1=0\,,\quad a_2=\psi_2=\frac{n_2\pi}{k_2}\,,\quad a_3=\psi_1=\frac{n_1\pi}{k_1}\,,
 \end{equation}
 with $n_i=0\,,\dots,k_i\,,\,i=1,2$.
 
 \noindent The highest dimensional branes correspond to \eqref{381} with $a_1,a_2,a_3\neq 0$ and as mentioned they correspond to consistent brane geometries for the UV and IR CFTs in \eqref{311'}. However they can not be identified as integrable brane geometries for the whole flow. To be more specific, if we consider the boundary values $(g_1,g_2)|_{\partial\S}=(r_1r_3,r_2r_3)$ with $r_i=h_if_ih_i^{-1}$ we find that the boundary contribution to the variation of \eqref{328} reads (see Appendix \ref{B3;})  
 \begin{equation}\label{ripa}
 \begin{split}
 \delta S|_{\partial\S}&=\frac{k_1}{2\pi}\int_{\partial\S}\Tr(\delta h_1h_1^{-1},\nabla_+g_1g_1^{-1}+\nabla_{r_3}g_1^{-1}\nabla_-g_1+(\mathbb{1}-D_{r_3})A_{1\tau}-\partial_{\tau}r_3r_3^{-1})\\
 &+\frac{k_2}{2\pi}\int_{\partial\S}\Tr(\delta h_2h_2^{-1},\nabla_+g_2g_2^{-1}+\nabla_{r_3}g_2^{-1}\nabla_-g_2+(\mathbb{1}-D_{r_3})A_{2\tau}-\partial_{\tau}r_3r_3^{-1})\\
 &+\frac{1}{2\pi}\int_{\partial\S}\Tr(\delta h_3h_3^{-1},(D_{r_3}-\mathbb{1})(k_1g_1^{-1}\nabla_-g_1+k_2g_2^{-1}\nabla_-g_2)\\
 &+(D_{r_3}-1)(k_1A_{1\tau}+k_2A_{2\tau})+(k_1+k_2)\partial_{\tau}r_3r_3^{-1})\,.
 \end{split}
 \end{equation}  
 Using now the constraints \eqref{c8}, this can be rewritten as
  \begin{equation}\label{ripa1}
  \begin{split}
  \delta S|_{\partial\S}&=\frac{k}{4\pi}\int_{\partial\S}\Tr(\delta h_1h_1^{-1},(\l^{-1}-1)(\mathcal{B}_+-D_{r_3}\mathcal{B}_-)+2s_1(\mathbb{1}-D_{r_3})A_{1\tau}-2s_1\partial_{\tau}r_3r_3^{-1})\\
  &+\frac{k}{4\pi}\int_{\partial\S}\Tr(\delta h_2h_2^{-1},(1-\l^{-1})(\mathcal{B}_+-D_{r_3}\mathcal{B}_-)+2s_2(\mathbb{1}-D_{r_3})A_{2\tau}-2s_2\partial_{\tau}r_3r_3^{-1})\\
  &+\frac{1}{2\pi}\int_{\partial\S}\Tr(\delta h_3h_3^{-1},
  (D_{r_3}-\mathbb{1})(k_1A_{1\tau}+k_2A_{2\tau})+(k_1+k_2)\partial_{\tau}r_3r_3^{-1})\,.
  \end{split}
  \end{equation}
 The vanishing of the boundary terms lead to the boundary conditions
\begin{equation}\label{gigi}
\begin{split}
((\l^{-1}+4s_1s_2)\mathbb{1}-&4s_1s_2D_{r_3})\mathcal{B}_+|_{\partial\S}= ((\l^{-1}+4s_1s_2)D_{L_3}-4s_1s_2\mathbb{1})\mathcal{B}_-|_{\partial\S}\,,\\
& \partial_{\tau}r_3r_3^{-1}|_{\partial\S}=(\mathbb{1}-D_{r_3})(s_1A_{1\tau}+s_2A_{2\tau})|_{\partial\S}\,.
\end{split}
\end{equation}  
 The field dependent gluing of the gauge fields $\mathcal{B}_{\pm}$ on the boundary, suggests that \eqref{gigi} could be derived with the method presented in section \ref{Ibc} by allowing the possibility to glue $T(\pi,0;z)$ to a gauge transformed reflected transport matrix. In this case the integrable boundary conditions take the form \cite{Driezen:2019ykp}
 \begin{equation}
 L_{\tau}(z)|_{\partial\S}=D_{r_3}L_{\tau}(z^{-1})+\partial_{\tau}r_3r_3^{-1}|_{\partial\S}\,,\quad r_3\in G\,.
 \end{equation}
 Using however the form of the Lax \eqref{11} it is straightforward to see that due to consistency $D^2_{r_3}=\mathbb{1}$, leading to $r_3=\mathbb{1}$ which is the case already studied.  
 \section{Generalized permutation branes in the integrable flows}\label{gpb}
 \subsection{Preliminaries on the generalized permutation branes}        
 In \cite{Fredenhagen:2005an} the authors defined a consistent brane configuration, involving a permutation automorphism, on a product of WZW models $G_{k_1}\times G_{k_2}$. Geometrically such branes, called generalized permutation branes, wrap the submanifold
\begin{equation}\label{251'}
{\mathcal{D}^{\pi}}_f=\left\{\left((h_1fh_2^{-1})^{k'_2},(h_2fh_1^{-1})^{k'_1}\right)=\left({v_1}^{k'_2},{v_2}^{k'_1}\right)|\forall\,h_1\,,\, h_2\in G\right\}\,,
\end{equation} 
 where $k'_i=k_i/k$ and $k=\text{gcd}(k_1,k_2)$. In general the dimensions of these branes are $2d_G-r_G$, but for $f=e$, where $e$ is the identity element of the group, is $d_G$ and the expression \eqref{251'} simplifies to
  \begin{equation}\label{252'}
 \mathcal{D}^{\pi}(e)=\left\{\left(v^{k'_2},v^{-k'_1}\right)|\,\forall\,v=h_1h_2^{-1}\in G\right\}\,.
 \end{equation}
Notice that in the case of equal levels the brane geometry \eqref{251'} reduce to that of the permutation branes \eqref{32}.

\noindent Furthermore, it was found that the restriction of the WZ three-form on \eqref{251'} satisfies the condition \eqref{21} with $\omega_{\text{WZ}}$ given by
 \begin{equation}\label{413}
 \begin{split}
 \omega_{\text{WZ}}(h_1,h_2)&=\frac{k'_1k'_2}{k}\left(\Tr(h_1^{-1}dh_1\wedge fh_2^{-1}dh_2f^{-1})+\Tr(h_2^{-1}dh_2\wedge fh_1^{-1}dh_1f^{-1})\right)\\
 &+k_1\sum_{i=1}^{k'_2-1}(k'_2-i)\Tr(v_1^iv_1^{-1}dv_1v_1^{-i}\wedge v_1^{-1}dv_1)\\
 &+k_2\sum_{i=1}^{k'_1-1}(k'_1-i)\Tr(v_2^iv_2^{-1}dv_2v_2^{-i}\wedge v_2^{-1}dv_2)\,.
 \end{split}
 \end{equation}
For topological reasons, discussed in \ref{Qc}, it is deduced in \cite{Fredenhagen:2005an} that $f=\exp(\pi i\l/\k)$ where $\l$ is an integral weight of $\mathfrak{g}$ and $\k=\text{lcm}(k_1,k_2)$. Thus, the number of stable generalized permutation branes coincides with the number of the ordinary branes in $G$ with level $\k$. 
  
\noindent Let us finally comment on the symmetries of the model in the presence of \eqref{251'}. It is invariant under the chiral transformation \eqref{313} in the bulk, which on the boundary reduces to
\begin{equation}\label{414'}
\begin{split}
&(g_1,g_2)|_{\partial\S}\mapsto (kg_1k^{-1},kg_2k^{-1})\,,
\end{split}
\end{equation}
in agreement with the boundary equations of motion found in \cite{Fredenhagen:2005an}
\begin{equation}\label{415}
k_1J_{1+}+k_2J_{2-}|_{\partial\S}=-k_1J_{1-}-k_2J_{2-}|_{\partial\S}\,,
\end{equation}
Thus, the generalized permutation branes preserve the $\hat{\mathfrak{g}}_{k_1+k_2}$ Kac-Moody algebra. Notice that the boundary conditions \eqref{415} do not preserve the Virassoro algebra associated with $\hat{\mathfrak{g}}_{k_1}\oplus\hat{\mathfrak{g}}_{k_2}$. However in \cite{Fredenhagen:2005an} and \cite{Fredenhagen:2009hx} the authors have proven that \eqref{251'} are consistent solutions of the DBI action, thus proving their conformal invariance.  
In the following sections we will investigate the possibility for such a brane geometry to belong in the family of integrable branes for the RG flows under study, specifically those with IR fixed points.
\subsection{Generalized permutation branes in the deformed models}
\subsubsection{GPB in the deformed coset space}
Let us now embed \eqref{251'} in \eqref{328}. In order to determine the boundary contribution in the variation of \eqref{328} first we have to specify the induced two-form on the brane worldvolume \eqref{251'}. It turns out that
\begin{align}\label{416}\nonumber
\omega_{k_1,k_2;\l}=&\frac{1}{4\pi}\Big(\omega_{WZ}(h_1,h_2)+\sum_{i=1}^{2}2k_i(1-\l)\Tr(dg_ig_i^{-1},\L^{-T}_{i,i+1}(s_ig_{i}^{-1}dg_{i}+s_{i+1}g_{i+1}^{-1}dg_{i+1}))|_{{\mathcal{D}^{\p}}_f}\\
&-2\l s_is_{i+1}k_i\Tr(dg_ig_i^{-1},\L_{i,i+1}^{-T}(D_{i+1}^T-\mathbb{1})g_i^{-1}dg_i)|_{{\mathcal{D}^{\p}}_f}\Big)\,,
\end{align}
where $\omega_{\text{WZ}}(h_1,h_2)$ is given in \eqref{413}. Then we find that
\begin{equation}\label{417}
\begin{split}
\delta S|_{\partial\S}&=\frac{k(\l^{-1}-1)}{4\pi}\int_{\partial\S}\Tr(\delta h_1h_1^{-1},((1-D^T_{v_1})^{-1}-(1-D_{v_2})^{-1})(\mathcal{B}_+-\mathcal{B}_-))\\
&-\frac{k(\l^{-1}-1)}{4\pi}\int_{\partial\S}\Tr(\delta h_2h_2^{-1},((1-D^T_{v_2})^{-1}-(1-D_{v_1})^{-1})(\mathcal{B}_+-\mathcal{B}_-))\,.
\end{split}
\end{equation}
It is straightforward to see that its vanishing leads to the conclusion that the brane geometry \eqref{251'} with the two-form \eqref{413} solve the boundary conditions \eqref{13}, thus consisting an integrable brane configuration.

\noindent The brane geometry \eqref{251'} preserves gauge invariance of model (III). To see this notice that the boundary conditions \eqref{251'} remain fixed under the gauge transformation \eqref{118}
\begin{equation}
(g_1,g_2)|_{\partial\S}\mapsto (Lg_1L^{-1},Lg_2L^{-1})|_{\partial\S}=((h'_1f{h'_2}^{-1})^{k'_2},(h'_2f{h'_1}^{-1})^{k'_1})\,,
\end{equation}
where $h'_1=Lh_1$ and $h'_2=Lh_2$. Setting thus,  $\delta_Vh_1h_1^{-1}=\delta_Vh_2h_2^{-1}$ in \eqref{417} we find that the boundary term in \eqref{557} vanishes, proving that model (III) preserves its gauge invariance under the presence of \eqref{251'}. Projecting the $(2d_G-r_G)$-dimensional branes in the deformed coset space we end up with $(d_G-r_G)$-dimensional integrable branes. For $G=SU(2)$ such branes are $D2$-branes.

\noindent According to the analysis of section \ref{Qc}, which may equally well be performed here, we conclude that the model (III) admits $\k=\text{lcm}(k_1,k_2)$ integrable branes, additionally to the $(k_1+1)(k_2+1)$ given in \eqref{B19''}, wrapped around ${\mathcal{D}^{\pi}}_f$.
\subsubsection{GPB in the deformed current algebra models}
For the sake of simplicity we will consider the lowest dimensional brane \eqref{252'} embedded in the target space of the sigma-model \eqref{320}. The boundary two-form, trivializing $H_{k_1,k_2;\l_1,\l_2}$ on \eqref{252'} is 
\begin{equation}\label{572}
\begin{split}
\omega_{k_1,k_2;\l_1,\l_2}=\frac{1}{4\pi}&\Big(\omega_{WZ}(h_1,h_2)+\sum_{i=1}^{2}\l_ik^{(i+1)}\Tr(dg_ig_i^{-1}\wedge\mathcal{O}^T_{i+1,i}g_{i+1}^{-1}dg_{i+1})|_{\mathcal{D}^{\pi}(e)}\\
&+\l_i\l_{i+1}k_i\Tr(dg_ig_i^{-1}\wedge \mathcal{O}^T_{i+1,i}D_{i+1}^Tg_i^{-1}dg_i)|_{\mathcal{D}^{\pi}(e)}\Big)\,,
\end{split}
\end{equation}
Varying the action and restricting our attention to the contribution from the boundary we find that (see Appendix \ref{B2;})
\begin{align}\label{419}
\delta S_{k_1,k_2;\l_1,\l_2}|_{\partial\S}&=\frac{\sqrt{k_1k_2}}{2\pi}\int_{\partial\S}\Tr(\delta h_1h_1^{-1},(1-D^T_{v})^{-1}((\l_1^{-1}-\l_0^{-1})A_{1+}+(\l_2^{-1}-\l_0)A_{2+}\nonumber\\\nonumber
&-(\l_1^{-1}-\l_0)A_{1-}-(\l_2^{-1}-\l_0^{-1})A_{2-})\\\nonumber
&+\frac{\sqrt{k_1k_2}}{2\pi}\int_{\partial\S}\Tr(\delta h_2h_2^{-1},(1-D_{v})^{-1}((\l_1^{-1}-\l_0^{-1})A_{1+}+(\l_2^{-1}-\l_0)A_{2+}\nonumber\\
&-(\l_1^{-1}-\l_0)A_{1-}-(\l_2^{-1}-\l_0^{-1})A_{2-})\,.
\end{align}
This indicates that the brane \eqref{252'} with the boundary two-form \eqref{572} is a solution of the boundary conditions
\begin{equation}\label{420}
(\l_1^{-1}-\l_0^{-1})A_{1+}+(\l_2^{-1}-\l_0)A_{2+}|_{\partial\S}=(\l_1^{-1}-\l_0)A_{1-}+(\l_2^{-1}-\l_0^{-1})A_{2-}|_{\partial\S}\,.
\end{equation}
Unfortunately, we were not able to derive \eqref{420} with the boundary monodromy method presented in the first section, thus we can not be sure if model (II) continues to be integrable in the presence of the brane \eqref{252'}. However, we will make a few observations regarding its presence in the RG flow under study and especially at its fixed points.

\noindent Consider the case where one of the deformation parameters freezes out, e.g. $(\l_1,\l_2)=(\l,0)$.\footnote{Notice that this RG flow can not be considered for the branes \eqref{22} studied in \ref{cft}, since they require $\l_1=\l_2$} This theory smoothly flows from a UV CFT, for $\l=0$, towards a CFT in the IR for $\l=\l_0$. That is
\begin{equation}\label{421}
G_{k_1}\times G_{k_2}\Longrightarrow G_{k_1}\times G_{k_2-k_1}\,.
\end{equation}
The IR CFT can be realized under the group redefinition $(g_1,g_2)\to(g'_1=g_2g_1,g'_2=g_2)$. It is straightforward to see that the boundary conditions \eqref{420} in the flow \eqref{421}, expressed in terms of the group elements are 
\begin{equation}\label{422}
(\l_0-\l)J_{1+}+\l_0^{-1}(J_{2+}+\l D_2J_{1+})|_{\partial\S}=(\l-\l_0^{-1})J_{2-}-\l_0(J_{1-}+\l D_1^TJ_{2-})|_{\partial\S}\,.
\end{equation}
In the UV limit, $\l=0$, \eqref{422} reduce to \eqref{415} as expected, in agreement with the symmetry \eqref{414'} preserved by the brane \eqref{252'}. Turning our attention to the IR limit, $\l=\l_0$, \eqref{422} becomes 
\begin{equation}\label{423}
k_1J'_{1+}+(k_2-k_1)J'_{2+}|_{\partial\S}=-k_1J'_{1-}-(k_2-k_1)J'_{2-}|_{\partial\S}\,,
\end{equation}
where we expressed the fields in terms of $(g'_1,g'_2)$.\footnote{We used the identities 
\begin{equation}
J'_{1+}=J_{2+}+D_2J_{1+}\,,\quad J'_{1-}=J_{1-}+D_1^TJ_{2-}\,.
\end{equation}} Notice that the currents appearing in \eqref{423} are the diagonal combination of the generators of the Kac-Moody algebra of the IR CFT. This agrees with the fact that the brane \eqref{252'} under the redefinition $(g_1,g_2)\to (g'_1,g'_2)$, transforms as
\begin{equation}\label{651}
{\mathcal{D}^{\pi}}_e=\left(v^{k'_2},v^{-k'_1}\right)\to\left(v^{k'_2-k'_1},v^{-k'_1}\right)\,,
\end{equation}
which corresponds to the lowest dimensional GPB embedded in the IR CFT $G_{k_1}\times G_{k_2-k_1}$, since $k=\text{gcd}(k_1,k_2)=\text{gcd}(k_1,k_2-k_1)$ for $k_2>k_1$. Let us point out that the above result holds only for the brane \eqref{252'} as the higher dimensional ones \eqref{251'}, under the previous field redefinition, transform as
\begin{equation}\label{653'}
{\mathcal{D}^{\pi}}_f\to\left\{\left({v_2}^{k'_2}{v_1}^{k'_1},{v_2}^{k'_2}\right)|\forall\,h_1\,,\, h_2\in G\right\}=\mathcal{M}_{f}\,,
\end{equation}
which do not belong in the class of the GPBs for the product $G_{k_1}\times G_{k_2-k_1}$.\footnote{Based on the arguments of \cite{Fredenhagen:2005an}, one can show that the number of stable $\mathcal{M}_f$ branes embedded in the IR CFT $G_{k_1}\times G_{k_2-k_1}$ is $\k=\text{lcm}(k_1,k_2)$. Thus, as the previously defined geometries the number of stable branes \eqref{251'} remains invariant along the whole deformation line \eqref{421}.}

\noindent Thus, we conclude that the brane configuration 
\begin{equation}\label{582}
\begin{split}
& \mathcal{D}^{\pi}(e)=\left\{\left(v^{k'_2},v^{-k'_1}\right)|\,\forall\,v=h_1h_2^{-1}\in G\right\}\,,\\
\omega_{k_1,k_2;\l}=&\omega_{WZ}(h_1,h_2)+\sqrt{k_1k_2}\l\Tr(dg_1g_1^{-1}\wedge g_2^{-1}dg_2)|_{\mathcal{D}^{\pi}(e)}\,,\\
\end{split}
\end{equation}
with $\omega_{\text{WZ}}(h_1,h_2)$ given in \eqref{413}, embedded in the RG flow \eqref{421} smoothly flows from a conformal brane in the UV, for $\l=0$, to a conformal brane in the IR for $\l=\l_0$. However its integrability preserving nature between the two CFT points remains open. We expect that the same results hold for the higher dimensional GPB branes given in \eqref{251'}. 
 
\noindent In the case where none of the deformation parameters is set to zero the model flows to the IR CFT given in \eqref{312} for $(\l_1,\l_2)=(\l_0,\l_0)$. The boundary conditions \eqref{420}, evaluated at the IR fixed point, become
\begin{equation}\label{426}
A_{2+}(\l_0,\l_0)|_{\partial\S}=A_{1-}(\l_0,\l_0)|_{\partial\S}\,,
\end{equation} 
which, as was mentioned before, relate the left-right generators of the current algebra symmetry of the IR CFT \eqref{312}. Thus the presence of the brane \eqref{252'} in the IR limit, breaks down the symmetry group \eqref{312} to the diagonal subgroup of \eqref{316}. Finally let us mention that in the case of equal levels $k_1=k_2$ the whole set up is reduced to the integrability preserving permutation branes \eqref{B20} embedded in model (I) \eqref{19a}.
\subsection{The lowest dimensional GPB for $G=SU(2)$ } 
\noindent The action describing the entire flow \eqref{421}, for a worldsheet with no boundaries, is
\begin{equation}\label{A1}
S_{k_1,k_2;\l}=S_{k_1}(g_1)+S_{k_2}(g_2)-\frac{\sqrt{k_1k_2}\l}{\pi}\int\Tr(J_{1+},J_{2-})\,.
\end{equation}
The induced metric and boundary two-form on the worldvolume of \eqref{252'} will be derived from the background fields of \eqref{A1} for $g_i=SU(2)$, $i=1,2$ and in the parametrization \eqref{A2}.
Doing so, we extract the background metric and $H$-field of \eqref{A1} but we do not present them here.

\noindent As has been noted in \cite{Fredenhagen:2005an} the parametrization \eqref{A2} proves to be convenient for the description of the brane \eqref{252'}. Being specific every point on the brane satisfies the relation
\begin{equation}\label{A3}
\psi_1=k'_2\psi\,,\quad\psi_2=-k'_1\psi\,,\quad\theta_1=\theta_2=\theta\,,\quad\phi_1=\phi_2=\phi.\, 
\end{equation}
where $g_i=g_i(\psi_i,\theta_i,\phi_i)$, $i=1,2$ and $v=v(\psi,\theta,\phi)$.
Thus, we find that the induced metric reads
\begin{equation}\label{A4}
\begin{split}
d\hat{s}^2&=f(k_1,k_2,\l)d\psi^2+g(k_1,k_2;\l)(d\theta^2+\sin^2(\theta)d\phi^2)\,,
\end{split}
\end{equation}
where
\begin{equation}\label{A5}
\begin{split}
&f(k_1,k_2,\l)=k'_1k'_2(k_1+k_2-2k_2\l\l_0)\,,\\
&g(k_1,k_2;\l)=(k_1-k_2\l\l_0)\sin^2(k'_2\psi)+k_2(1-\l\l_0)\sin^2(k'_1\psi)\\
&\hspace{20mm}+k_2\l\l_0\sin^2((k'_1-k'_2)\psi)\,.
\end{split}
\end{equation}
From \eqref{582} and using the form of $\eqref{413}$ found in \cite{Fredenhagen:2005an} we find that the boundary two- form is
\begin{equation}\label{A6}
\omega_{k_1,k_2;\l}=h(k_1,k_2,\l)\sin(\theta)d\theta\wedge d\phi\,,
\end{equation}
where
\begin{equation}\label{A7}
\begin{split}
h(k_1,k_2,\l)&=k_2\sin(2k'_1\psi)-k_1\sin(2k'_2\psi)\\
&-4k_2\l_0\l\sin(k'_2\psi)\sin(k'_1\psi)\sin((k'_2-k'_1)\psi)\,.
\end{split}
\end{equation}
It is straightforward to see that for $\l=0$ the induced fields $\eqref{A4}$ and $\eqref{A6}$ become those found in \cite{Fredenhagen:2005an}. Turning our attention to the IR limit $\l=\l_0=\sqrt{k_1/k_2}$ we find what we expected from the analysis in the previous section, i.e. the fields on the brane are the ones found in the UV limit under the redefinition $k_2\to k_2-k_1$ 
\begin{align}
&d\hat{s}^2=k'_1(k'_2-k'_1)k_2d\psi^2+(k_1\sin^2((k'_2-k'_1)\psi)-(k_2-k_1)\sin^2k'_1\psi)(d\theta^2+\sin(\theta)^2d\phi^2)\,,\nonumber\\
&\omega_{k_1,k_2,\l_0}=(k_1\sin(2(k'_2-k'_1)\psi)-(k_2-k_1)\sin(2k'_1\psi))\sin(\theta)d\theta\wedge d\phi\,.
\end{align}
Notice that in the case of equal levels $k_1=k_2$ the whole set up is reduced to the lowest dimensional integrability preserving permutation brane, ${\mathcal{C}^{\pi}}_{f,f^{-1}}$, embedded in the RG flow described by \eqref{16'} for $(\l,\l_2)=(\l,0)$. It is straightforward to see that the metric \eqref{A4} reduces to the metric of an $S^3$-sphere with a deformed radius embedded in the diagonal and symmetric way in the target space of the corresponding sigma-model  
\begin{equation}
ds^2=k(1-\l)(d\psi^2+\sin^2\psi(d\theta^2+\sin^2\theta d\phi^2))\,.
\end{equation}
In this case the two-form \eqref{A6} vanishes, in agreement with \cite{Figueroa-OFarrill:2000gfl} \cite{Sarkissian:2003yw}. 
\section{Conclusions}
In this paper we studied integrable brane configurations embedded in three types of generalized $\l$-deformed models, denoted as model (I), (II) and (III). The model (I) and (II) correspond to the deformation of two WZW models based on the product group, $G_{k}\times G_k$ and $G_{k_1}\times G_{k_2}$ respectively. The model (III) corresponds to the deformation of a diagonal coset space CFT of the form $G_{k_1}\times G_{k_2}/G_{k_1+k_2}$. Model (I) flows to a strongly coupled theory, while (II) and (III) to exact CFTs in the far IR.

\noindent To obtain the integrable boundary conditions along their RG flows, we constructed the simplest forms of boundary monodromy matrices and demanded, in each case, that they generate conserved charges in the presence of boundaries. Using two different definitions of the reflection transformation we found that the models (I) and (II) admit two kinds of integrable conditions, relating the components of the same and of different gauge fields. In the case of (I), both types are consistent with the vanishing of the momentum flow from the boundary, and have a nice geometric interpretation as a product of G-conjugacy classes, $\mathcal{C}_{f_1f_2}$, and twisted conjugacy classes, ${\mathcal{C}^{\pi}}_{f_1f_2}$. We saw that the conjugacy classes require the couplings to be equal, $\l_1=\l_2$, while for the twisted ones this is not a requirement. Model (II) admits one set of integrable boundary conditions consistent with the vanishing of the momentum flow, which correspond to branes wrapping around $\mathcal{C}_{f_1f_2}$. The other set, even though it preserves integrability of the model, does not have a geometric interpretation in terms of Dirichlet and Neumann boundary conditions. Finally, model (III) admits integrable brane configurations that are also described by a product of $G$-conjugacy classes.  

\noindent After the identification of the integrable boundary conditions and their geometric interpretation as D-branes, we proceeded with the study of their stability points and their geometry. As in \cite{Driezen:2018glg} we found that all occurrences of the deformation in the quantization condition \eqref{q} cancel out, enforcing the D-branes to sit at localized positions. This result, in the case of models (II) and (III), indicates that the number of stable branes in the UV and IR fixed points of their RG flows remains invariant, despite the loss of bulk degrees of freedom dictated by Zamolodchikov's C-theorem. Subsequently, we proceeded with the identification of the surviving symmetries of the latter CFTs. Being specific, the model (II) flows from the CFT \eqref{311} to the CFT \eqref{312}. The presence of the brane $\mathcal{C}_{f_1f_2}$ along its deformation line preserves in the UV the diagonal subgroup of each copy of the chiral symmetry \eqref{313}. In the IR it can be realized as the gauge fixed version of a brane geometry that respects its gauge symmetry \eqref{38} and the diagonal subgroup of \eqref{39}. For the model (III) we found that the integrable branes preserve its gauge invariance along the whole deformation line. Additionally, for $G=SU(2)$ and in the two CFT limits, we saw that they can be identified with the $0,1-$dimensional Cardy branes for the diagonal coset spaces $SU(2)_{k}\times SU(2)_{l}/SU(2)_{k+l}$.   
   
\noindent Finally, we considered the newest class, at least to our knowledge, of non factorizing D-branes known as generalized permutation branes, ${\mathcal{D}^{\pi}}_f$. We saw that they can be thought of as additional integrable brane geometries for model (III), which also preserve its gauge symmetry. On the contrary we were not able to identify them as integrable brane geometries for model (II), however we saw that they interpolate between two conformal brane configurations in the UV and IR fixed points of its RG flow. As the previously defined geometries, their quantization condition remains invariant when we turn on the deformation leading to an invariant brane spectrum between the two CFT limits of (II) and (III). Finally, we considered the lowest dimensional GPB embedded in the deformed $SU(2)_{k_1}\times SU(2)_{k_2}$ CFT and we extracted its induced fields.

\noindent There are several directions for future study: One, for example, could try to construct a more exotic generalized transport matrix \eqref{3} in order to derive the boundary conditions \eqref{420} as conditions that preserve the bulk integrability of model (II) and subsequently derive the  infinite tower of conserved local and non local higher spin charges. Furthermore, it would be interesting to find integrable boundary conditions for the models recently constructed in \cite{Georgiou:2018hpd,Georgiou:2018gpe,Georgiou:2020wwg,Georgiou:2021pbd} which admit Lax pairs that do not have a simple spectral dependence, and apply the sigma-model approach in order to give a geometric picture as D-branes. Lastly, one might want to consider the boundary conditions we studied on the upper half plane as was done in \cite{Sfetsos:2021pcs} in the case of the single $\l$-deformed model. 
\section*{Acknowledgement}
I thank G. Georgiou and K. Siampos for usefull discussions on the topic of D-branes. I also appreciate the fruitful and very interesting comments of S. Driezen and D. Thompson on this work. Finally I am grateful to K. Sfetsos for suggesting this topic and for a careful reading of the manuscript. Part of this work was funded by the State Scholarships Foundation (I.K.Y). 
\appendix
\section{Integrable branes in the $\l$-model}\label{A}
Here we will show in detail, using the approach of section \ref{Ibc}, that the brane geometry
\begin{equation}\label{A2''''}
{\mathcal{C}^{\omega}}_f=\{\omega(h)fh^{-1}|\forall h\in G\}\,.
\end{equation}
corresponds to the integrable boundary conditions found in \cite{Driezen:2018glg}\cite{Driezen:2019ykp}.

\noindent The effective $\l$-deformed group $G$ and symmetric space $G/H$ action can be put in the following compact form
\begin{equation}\label{A11}
S_{k;\l}=S_{k}(g)-\frac{k}{\pi}\int_{\S}\Tr(\partial_+gg^{-1},(\mathcal{P}-D^T)^{-1}g^{-1}\partial_-g)\,,
\end{equation}
where the operator $\mathcal{P}$ is defined as
\begin{equation}\label{A1'}
\mathcal{P}=\begin{cases}
\l^{-1},\quad\text{group space}\,,\\
\mathcal{P}^{(0)}+\l^{-1}\mathcal{P}^{(1)},\quad\text{symmetric space}\,.
\end{cases}
\end{equation}
In the latter case $\mathcal{P}^{(i)}$ are the projectors along the $\mathbb{Z}_2$-decomposition of the Lie algebra $\mathfrak{g}=\mathfrak{g}^{(0)}\oplus\mathfrak{g}^{(1)}$.

\noindent In the presence of boundaries the action \eqref{A11} is modified as
\begin{equation}\label{A2''''''}
S_{k;\l}=\int_{\S}L_{k;\l}+\int_{M'}H_{k;\l}-\int_{D}\omega_{k;\l}\,,
\end{equation}
where\footnote{Our worldsheet conventions are
\begin{equation}\label{lol}
\eta_{\tau\tau}=-\eta_{\s\s}=\eta^{\tau\tau}=-\eta^{\s\s}=1\,,\quad \e_{\tau\s}=-\e^{\tau\s}=-\e_{\s\tau}=\e^{\s\tau}=-1\,.
\end{equation}
Lightcone coordinates are defined as $\s_{\pm}=\tau\pm\s$.}
\begin{equation}\label{A2'''}
\begin{split}
&L_{k,\l}=-\frac{k}{8\pi}\Tr(g^{-1}\partial_{\m}g,g^{-1}\partial^{\m}g)-\frac{k}{4\pi}\Tr(\partial^{\m}gg^{-1},(\mathcal{P}-D^T)^{-1}g^{-1}\partial_{\m}g)\,,\\
&H_{k,\l}(g)=\frac{k}{4\pi}(H_{WZ}(g)+d\Tr(dgg^{-1}\wedge(\mathcal{P}-D^T)^{-1}g^{-1}dg))\,.
\end{split}
\end{equation}
The three-dimensional space $M'$ is defined such that $\partial M'=\S\cup D$ and
\begin{equation}\label{268}
H_{k,\l}|_{\text{brane}}=d\omega_{k,\l}\,.
\end{equation}
Using the brane geometry \eqref{A2''''} and the equation \eqref{268} we find that 
\begin{equation}\label{269}
\omega_{k,\l}(h)=\frac{k}{4\pi}(\omega_{WZ}(h)+\Tr(dgg^{-1}\wedge(\mathcal{P}-D^T)^{-1}g^{-1}dg)|_{{\mathcal{C}^{\omega}}_f})\,,
\end{equation}
where \cite{Stanciu:2000fz}
\begin{equation}
\omega_{\text{WZ}}(f)=\tr(h^{-1}dh\wedge\Omega ^Tf h^{-1}dhf^{-1})\,,
\end{equation}
with $\Omega$ satisfying the condition $\Omega^T=\Omega^{-1}$.\footnote{To see this notice that in order for $H_{WZ}(g)|_{{\mathcal{C}^{\omega}}_f}$ to equal $\omega_{WZ}(f)$ the following relation must hold
\begin{equation}
\Tr(t^A,[t^B,t^C])=\Tr(\Omega(t^A),[\Omega(t^B),\Omega(t^C)])=\Tr(\Omega(t^A),\Omega[t^B,t^C])\,.
\end{equation}}
Now we are in position to derive the boundary conditions which correspond to the brane \eqref{A2''''}. We begin with the variation of the first term in \eqref{A2''''''} where we restrict ourselves in the boundary terms while the bulk terms will be omitted. Doing so we find that 
\begin{equation}\label{A3'}
\begin{split}
\int\delta L_{k;\l}|_{\partial\S}&=\frac{k}{4\pi}\int_{\partial\S}(\Tr(g^{-1}\delta g,g^{-1}\partial_{\s}g)+\Tr(\delta gg^{-1},(\mathcal{P}-D^T)^{-1}g^{-1}\partial_{\s}g)\\
&+\Tr(g^{-1}\delta g, (\mathcal{P}-D)^{-1}\partial_{\s}gg^{-1})\\
&=\frac{k}{4\pi}\int_{\partial\S}(\Tr(g^{-1}\delta g,g^{-1}\partial_{\s}g)-\Tr(\delta gg^{-1},A^L_{\s})+\Tr(g^{-1}\delta g, A^R_{\s}))\,,
\end{split}
\end{equation}
where for convenience we have defined
\begin{equation}\label{272}
\begin{split}
A^L_{\m}=-(\mathcal{P}-D^T)^{-1}g^{-1}\partial_{\m}g,\quad A^R_{\m}=(\mathcal{P}-D)^{-1}\partial_{\m}gg^{-1}\,,\quad \m=\tau,\s\,.
\end{split}
\end{equation}
Using \eqref{A2''''} one can see that on the boundary, arbitrary variations of $g$ can be written as  
\begin{equation}\label{A0}
\begin{split}
&\delta gg^{-1}|_{\partial\S}=(\Omega-D)\delta hh^{-1}\,,\\
&g^{-1}\delta g|_{\partial\S}=(D^T\Omega-\mathbb{1})\delta hh^{-1}\,.
\end{split}
\end{equation}
Plugging \eqref{A0} in \eqref{A3'} we find that 
\begin{equation}
\begin{split}\label{A8}
\int\delta L_{k;\l}|_{\partial\S}=\frac{k}{4\pi}\int_{\partial\S}\Tr(\delta hh^{-1},\Omega^T\partial_{\s}gg^{-1}-g^{-1}\partial_{\s}g-(\Omega^T- D^T)A^L_{\s}+(\Omega^T D-\mathbb{1})A^R_{\s})\,.
\end{split}
\end{equation}
Similarly, the variation of the boundary two-form \eqref{269} reads \footnote{We used the fact that $\partial\S=-\partial D$ and the relation
\begin{equation}\label{vfv}
(D^T\Omega-\Omega^TD)\partial_{\tau}hh^{-1}=\Omega^T g^{-1}\partial_{\tau}g+\partial_{\tau}gg^{-1}\,.
\end{equation} } 
\begin{equation}\label{A9}
\begin{split}
\int\delta\omega_{k;\l}|_{\partial D} =-\frac{k}{4\pi}\int_{\partial\S}\Tr(\delta hh^{-1},\Omega^T\partial_{\tau}gg^{-1}+(\Omega^T D-\mathbb{1})A^R_{\tau}+g^{-1}\partial_{\tau}g+(\Omega^T-D^T)A^L_{\tau})\,.
\end{split}
\end{equation}
Combining \eqref{A8},\eqref{A9} we find that the total boundary contribution is  
\begin{equation}\label{A10'}
\delta S|_{\partial\S}=\frac{k}{2\pi}\int_{\partial\S}\tr(\delta hh^{-1},\Omega^T\nabla_+gg^{-1}+g^{-1}\nabla_-g+(\Omega^T-\mathbb{1})(A_++A_-))\,.
\end{equation}
Specializing to the case of the $\l$-deformed $G$-model this becomes \footnote{We used the constraint equations for the $\l$ $G$-model \begin{equation}\label{r}
	\nabla_+gg^{-1}=(\l^{-1}-1)A_+,\quad g^{-1}\nabla_-g=-(\l^{-1}-1)A_-
	\end{equation}}
\begin{equation}\label{A10'''}
\delta S|_{\partial\S}=\frac{k}{2\pi}\int_{\partial\S}\tr(\delta hh^{-1},(\l^{-1}\Omega^T-\mathbb{1})A_+-(\l^{-1}\mathbb{1}-\Omega^T)A_-)\,.
\end{equation}
Thus the brane configuration $\eqref{A2''''}$ and $\eqref{269}$ is a solution of the boundary conditions
\begin{equation}\label{A10''}
(\l^{-1}\mathbb{1}-\Omega)A_+|_{\partial\S}=(\l^{-1}\Omega-\mathbb{1})A_-|_{\partial\S}\,,
\end{equation}
where we used also the fact that $\Omega^T=\Omega^{-1}$. We know from \cite{Driezen:2018glg} that in order for \eqref{A2''''} to be an integrable brane geometry $\Omega$ should be an involutive automorphism. In this case \eqref{A10''} reduces to the integrable boundary conditions, found in the same work,
\begin{equation}\label{A20}
A_+|_{\partial\S}=\Omega A_{-}|_{\partial\S}\,,\quad \Omega^2=\mathbb{1}\,.
\end{equation}
Turning our attention now to the $\l$-deformed $G/H$-model we decompose $\delta hh^{-1}$ in its subgroup and coset components denoted with the indices $(0)$ and $(1)$ respectively. We also assume that $\Omega$ respects the $\mathbb{Z}_2$-decomposition of $\mathfrak{g}$. Doing so we find that \eqref{A10'} becomes \footnote{We used the constraint equations for the $\l$ $G/H$-model
\begin{equation}\label{337}
\begin{split}
& (\nabla_+gg^{-1})^{(0)}=0,\quad (g^{-1}\nabla g)^{(0)}=0\,,\\
&(\nabla_+gg^{-1})^{(1)}=(\l^{-1}-1)A^{(1)}_+,\quad (g^{-1}\nabla_-g)^{(1)}=-(\l^{-1}-1)A^{(1)}_-\,.
\end{split}
\end{equation}}
\begin{equation}\label{A10''''}
\begin{split}
\delta S|_{\partial\S}&=\frac{k}{2\pi}\int_{\partial\S}\tr({\delta hh^{-1}}^{},(\Omega^T-1)A_{\tau})_{(0)}\\
&+\tr(\delta hh^{-1},(\l^{-1}\Omega^T-\mathbb{1})A_+-(\l^{-1}\mathbb{1}-\Omega^T)A_-)_{(1)}\,.
\end{split}
\end{equation}
The vanishing of \eqref{A10''''} leads to the boundary conditions
$\Omega(\mathfrak{g}^{(0)})=\mathbb{1}$ unless $A^{(0)}_{\tau}|_{\partial\S}=0$ and $A^{(1)}_+|_{\partial\S}=\Omega A^{(1)}_-|_{\partial\S}$ with $\Omega^2(\mathfrak{g}^{(1)})=\mathbb{1}$. These are exactly the integrable boundary conditions found in \cite{Driezen:2019ykp} for $\mathcal{W}=\mathbb{1}$.
\section{Derivation of the boundary terms for model (I)-(III)}\label{B''}
In this section we will present the technical details for the derivation of the boundary conditions for the integrable branes of models (I), (II) and (III). 
\subsection{Model (I)}\label{bconI}
We start with the variation of the first term in \eqref{20}. As before we will concentrate ourselves only on boundary terms. For further simplification we will omit the label $(k;\l_1,\l_2)$ and as before we will define,
\begin{equation}\label{1a'}
\begin{split}
&A^R_{1\m}=\l_1(\mathbb{1}-\l_1\l_2D_1D_2)^{-1}(\partial_{\m}g_1g_1^{-1}+\l_2D_1\partial_{\m}g_2g_2^{-1})\,,\\
&A^{L}_{1\m}=-\l_1(\mathbb{1}-\l_1\l_2D_2^TD_1^T)^{-1}(g^{-1}_2\partial_{\m}g_{2}+\l_2D_2^Tg^{-1}_1\partial_{\m}g_1)\,,\\
&A^R_{2\m}=\l_2(\mathbb{1}-\l_1\l_2D_2D_1)^{-1}(\partial_{\m}g_2g_2^{-1}+\l_1D_2\partial_{\m}g_1g_1^{-1})\,,\\
&A^L_{2\m}=-\l_2(\mathbb{1}-\l_1\l_2D_1^TD_2^T)^{-1}(g^{-1}_{1}\partial_{\m}g_1+\l_1D_1^Tg^{-1}_2\partial_{\m}g_2)\,,
\end{split}
\end{equation}
with $\m=\tau,\s$. Doing so we find 
\begin{equation}\label{B1}
\begin{split}
\int\delta L|_{\partial\S}&=\frac{k}{4\pi}\int_{\partial\S}\sum_{i=1}^{2}\Tr(g_i^{-1}\delta g_i,g_i^{-1}\partial_{\s}g_i)-\Tr(\delta g_ig_i^{-1},A^L_{i\s})+\Tr(g_i\delta g_i,A^R_{i+1\s})\,.
\end{split}
\end{equation} 
Using \eqref{22} we find that arbitrary variations of the group elements on the boundary can be written as
\begin{equation}\label{B2'}
\begin{split}
&g_i^{-1}\delta g_i=(D_i^T-1)\delta h_ih_i^{-1}\,,\\
&\delta g_ig_i^{-1}=(1-D_i)\delta h_ih_i^{-1}\,,\quad i=1,2\,.
\end{split}
\end{equation}
Substituting \eqref{B2'} in \eqref{B1} we find
\begin{equation}\label{B3}
\begin{split}
\int\delta L|_{\partial\S}
&=\frac{k}{4\pi}\int_{\partial\S}\sum_{i=1}^{2}\Tr(\delta h_ih_i^{-1},\partial_{\s}g_ig_i^{-1}-g^{-1}_i\partial_{\s}g_i-(1-D_i^T)A^L_{i\s}+(D_i-1)A^R_{i+1\s})\,.
\end{split}
\end{equation}
Similarly, the variation of the boundary two-form \eqref{530} gives
\begin{equation}\label{B4}
\begin{split}
\int\delta\omega|_{\partial D}&=-\frac{k}{4\pi}\int_{\partial\S}\sum_{i=1}^{2}\Tr(\delta h_ih_i^{-1},\partial_{\tau}g_ig_i^{-1}+g_i^{-1}\partial_{\tau}g_i+(D_i-1)A^R_{i+1\tau}+(1-D_i^T)A^L_{i\tau})\,.
\end{split}
\end{equation} 
Combining \eqref{B3} and \eqref{B4} we find that the total boundary contribution reads
\begin{equation}\label{B5}
\begin{split}
\delta S|_{\partial\S}
&=\frac{k}{2\pi}\int_{\partial\S}\sum_{i=1}^2\Tr(\delta h_ih_i^{-1},\partial_{+}g_ig_i^{-1}+g^{-1}_i\partial_{-}g_i+(1-D_i^T)A_{i-}+(D_i-1)A_{i+1+})\\
&=\frac{k}{2\pi}\int_{\partial\S}\sum_{i=1}^{2}\Tr(\delta h_ih_i^{-1},\nabla_+g_ig_i^{-1}+g_i^{-1}\nabla_-g_i+(A_{i+}-A_{i+1+})+(A_{i-}-A_{i+1-}))\,.
\end{split}
\end{equation}
To pass from the first to the second equality we used the definition of the covariant derivative $\nabla_{\m}g_i=\partial_{\m}g_i-A_{i\m}g_i+g_iA_{i+1\m}$ \cite{Georgiou:2017jfi}.

\noindent Turning our attention to the brane \eqref{32} we find that arbitrary variations of the group elements on the boundary can be written as
\begin{equation}\label{B6}
\begin{split}
&g_i^{-1}\delta g_i=D_i^T\delta h_ih_i^{-1}-\delta h_{i+1}h_{i+1}^{-1}\,,\\
&\delta g_ig_i^{-1}=\delta h_ih_i^{-1}-D_i\delta h_{i+1}h_{i+1}^{-1}\,,\quad i=1,2\,.
\end{split}
\end{equation}
Following the same steps as before we find
\begin{equation}\label{B7}
\begin{split}
\int_{\S}\delta L&=\frac{k}{4\pi}\int_{\partial\S}\sum_{i=1}^{2}\Tr(\delta h_ih_i^{-1},\partial_{\s}g_ig_i^{-1}-A^R_{i\s}-D_iA^R_{i+1\s}-g_{i+1}^{-1}\partial_{\s}g_{i+1}+D_{i+1}^TA^L_{i+1\s}-A^L_{i\s})\,,
\end{split}
\end{equation} 
and
\begin{equation}\label{B8}
\begin{split}
\int_{D}\delta\omega&=-\frac{k}{4\pi}\int_{\partial\S}\sum_{i=1}^{2}\Tr(\delta h_ih_i^{-1},\partial_{\tau}g_ig_i^{-1}-A^R_{i\tau}-D_iA^R_{i+1\tau}+g_{i+1}^{-1}\partial_{\tau}g_{i+1}-D_{i+1}^TA^L_{i+1\tau}+A^L_{i\tau})\,,
\end{split}
\end{equation}
where the two-form is given in \eqref{B20}. Combining \eqref{B7} and \eqref{B8} we find that the total boundary contribution in the variation of \eqref{20} is  
\begin{align}\label{B9}\nonumber
\delta S_{\partial\S}&=\frac{k}{2\pi}\int_{\partial\S}\sum_{i=1}^2\Tr(\delta h_ih_i^{-1},\partial_+g_ig_i^{-1}-A_{i+}+D_iA_{i+1+}+g_{i+1}^{-1}\partial_-g_{i+1}-D_{i+1}^TA_{i+1-}+A_{i-})\\
&=\frac{k}{2\pi}\int_{\partial\S}\sum_{i=1}^2\Tr(\delta h_ih_i^{-1},\nabla_+g_ig_i^{-1}+g_{i+1}^{-1}\nabla_-g_{i+1})\,.
\end{align}
 \subsection{Model (II)}\label{B2;}
Since the boundary conditions \eqref{B27} correspond to the brane \eqref{22}, we skip the technical details as they are similar to \eqref{B1}-\eqref{B5}. We will show however that the brane geometry \eqref{252'} with the boundary two form \eqref{572} corresponds to the boundary conditions \eqref{420}. 

\noindent Varying the first term in \eqref{320} we find     
\begin{equation}\label{B10}
\begin{split}
\int\delta L|_{\partial\S}&=\sum_{i=1}^2\frac{k_i}{4\pi}\int_{\partial\S}\Tr(g_i^{-1}\delta g_i,g_i^{-1}\partial_{\s}g_i)-\Tr(\delta g_ig_i^{-1},A^L_{i\s})+\Tr(g_i\delta g_i,A^R_{i+1\s})\,,
\end{split}
\end{equation} 
where
\begin{equation}\label{1a''}
\begin{split}
&A^R_{1\m}=\l_1(\mathbb{1}-\l_1\l_2D_1D_2)^{-1}(\l_0\partial_{\m}g_1g_1^{-1}+\l_2D_1\partial_{\m}g_2g_2^{-1})\,,\\
&A^{L}_{1\m}=-\l_1(\mathbb{1}-\l_1\l_2D_2^TD_1^T)^{-1}(\l_0^{-1}g^{-1}_2\partial_{\m}g_{2}+\l_2D_2^Tg^{-1}_1\partial_{\m}g_1)\,,\\
&A^R_{2\m}=\l_2(\mathbb{1}-\l_1\l_2D_2D_1)^{-1}(\l_0^{-1}\partial_{\m}g_2g_2^{-1}+\l_1D_2\partial_{\m}g_1g_1^{-1})\,,\\
&A^L_{2\m}=-\l_2(\mathbb{1}-\l_1\l_2D_1^TD_2^T)^{-1}(\l_0g^{-1}_{1}\partial_{\m}g_1+\l_1D_1^Tg^{-1}_2\partial_{\m}g_2)\,.
\end{split}
\end{equation}
Arbitrary variations of the group elements on the boundary can be written as \cite{Fredenhagen:2005an}
\begin{equation}\label{B11}
\begin{split}
&g_1^{-1}\delta g_1|_{\partial\S}=\frac{1-D_1^T}{1-D_{v}^T}(D^T_{v}\delta h_1h_1^{-1}-\delta h_{2}h_{2}^{-1})\,,\\
&g_2^{-1}\delta g_2|_{\partial\S}=\frac{1-D_2^T}{1-D_{v}}(D_{v}\delta h_2h_2^{-1}-\delta h_{1}h_{1}^{-1})\,.
\end{split}
\end{equation} 
Substituting \eqref{B11} in \eqref{B10} we find
\begin{align}\label{B13}\nonumber
\int\delta L|_{\partial\S}
&=\frac{1}{4\pi}\int_{\partial\S}\sum_{i=1}^{2}\Tr((D_{r_i}-1)^{-1}\delta h_ih_i^{-1},k_i(\partial_{\s}g_ig_i^{-1}-g^{-1}_i\partial_{\s}g_i-(1-D_i^T)A^L_{i\s}\\\nonumber
&+(D_i-1)A^R_{i+1\s})
+k_{i+1}(\partial_{\s}g_{i+1}g_{i+1}^{-1}+g^{-1}_{i+1}\partial_{\s}g_{i+1}-(1-D_{i+1}^T)A^L_{i+1\s}\\
&+(D_{i+1}-1)A^R_{i\s}))\,.
\end{align}
where we defined $r_1=r_2^{-1}=v$. The variation of the WZ boundary two-form \eqref{413} has been computed in \cite{Fredenhagen:2005an}. Thus the variation of \eqref{572} reads  
\begin{align}\label{B14}\nonumber
\int\delta\omega|_{\partial D}
&=-\frac{1}{4\pi}\int_{\partial\S}\sum_{i=1}^{2}\Tr((D_{r_i}-1)^{-1}\delta h_ih_i^{-1},k_i(\partial_{\tau}g_ig_i^{-1}+g^{-1}_i\partial_{\tau}g_i+(1-D_i^T)A^L_{i\tau}\\\nonumber
&+(D_i-1)A^R_{i+1\tau})
+k_{i+1}(\partial_{\tau}g_{i+1}g_{i+1}^{-1}+g^{-1}_{i+1}\partial_{\tau}g_{i+1}-(1-D_{i+1}^T)A^L_{i+1\tau}\\
&+(D_{i+1}-1)A^R_{i\tau}))\,.
\end{align} 
Combining \eqref{B13} and \eqref{B14} we find that the total contribution from the boundary in the variation of \eqref{320} is 
\begin{align}\label{B15}\nonumber
\delta S_{\partial\S}
=\frac{1}{2\pi}\int_{\partial\S}&\sum_{i=1}^{2}\Tr((D_{r_i}-1)^{-1}\delta h_ih_i^{-1},k_i(\nabla_+g_ig_i^{-1}+g_i^{-1}\nabla_-g_i+A_{i+}-A_{i+1+}+\\\nonumber
&A_{i-}
-A_{i+1-})+k_{i+1}(\nabla_+g_{i+1}g_{i+1}^{-1}+g_{i+1}^{-1}\nabla_-g_{i+1}+A_{i+1+}-A_{i+}\\
&+A_{i+1-}-A_{i-}))\,.
\end{align} 
\subsection{Model (III)}\label{B3;}
Here we will skip the intermediate details and we will just present the necessary results for the derivation of the boundary conditions \eqref{ripa}, \eqref{417} 
\subsubsection{Boundary conditions \eqref{ripa} }
\begin{equation}
\begin{split}
&\delta g_ig_i^{-1}|_{\partial\S}=(1-D_iD_{r_3}^T)\delta h_ih_i^{-1}+(D_iD_{r_3}^T-D_i)\delta h_3h_3^{-1}\\
&g_i^{-1}\delta g_i|_{\partial\S}=(D^T_i--D_{r_3}^T)\delta h_ih_i^{-1}+(D_{r_3}^T-1)\delta h_3h_3^{-1}\,,\quad i=1,2\,,
\end{split}
\end{equation}
\begin{align}\label{C10}\nonumber
\int\delta L|_{\partial\S}&=\sum_{i=1}^2\frac{k_i}{4\pi}\int_{\partial\S}\Tr(\delta h_ih_i^{-1},\partial_{\s}g_ig_i^{-1}-D_{r_3}g_i\partial_{\s}g_i-(1-D_{r_3}D^T_i)A^L_{i\s}+(D_i-D_{r_3})A^{R}_{i\s})\\
&+\sum_{i=1}^{2}\frac{k_i}{4\pi}\int_{\partial\S}\Tr(\delta h_3h_3^{-1},(D_3-1)g_i^{-1}\partial_{\s}g_i-(D_{r_3}D^T_i-1)A^{L}_{i\s}+(D_{r_3}-1)A^R_{i\s})\,,
\end{align} 
\begin{align}\label{C11}\nonumber
\int\delta \omega|_{\partial D}&=-\sum_{i=1}^2\frac{k_i}{4\pi}\int_{\partial\S}\Tr(\delta h_ih_i^{-1},\partial_{\tau}g_ig_i^{-1}+D_{r_3}g_i\partial_{\tau}g_i+(1-D_{r_3}D^T_i)A^L_{i\tau}+(D_i-D_{r_3})A^{R}_{i\tau}\\\nonumber
&-2\partial_{\tau}r_3r_3^{-1})-\sum_{i=1}^{2}\frac{k_i}{4\pi}\int_{\partial\S}\Tr(\delta h_3h_3^{-1},(D_3-1)g_i^{-1}\partial_{\tau}g_i+(D_{r_3}D^T_i-1)A^{L}_{i\tau}\\
&+(D_{r_3}-1)A^R_{i\tau}+2\partial_{\tau}r_3r_3^{-1})\,.
\end{align}
\subsubsection{Boundary conditions \eqref{417}}
 \begin{equation}\label{B11'}
 \begin{split}
 &g_1^{-1}\delta g_1|_{\partial\S}=\frac{1-D_1^T}{1-D_{v_1}^T}(D^T_{v_1}\delta h_1h_1^{-1}-\delta h_{2}h_{2}^{-1})\,,\\
 &g_2^{-1}\delta g_2|_{\partial\S}=\frac{1-D_2^T}{1-D_{v_2}^T}(D^T_{v_2}\delta h_2h_2^{-1}-\delta h_{1}h_{1}^{-1})\,.
 \end{split}
 \end{equation}
 \begin{align}\nonumber
 \int\delta L|_{\partial\S}=\frac{1}{4\pi}\sum_{i=1}^{2}\int_{\partial\S}&\Tr(\delta h_ih_i^{-1},k_i(1-D^T_{v_i})^{-1}(\partial_{\s}g_ig_i^{-1}-g_i^{-1}\partial_{\s}g_i-(1-D_i^T)A^L_{i\s}\\\nonumber
 &+(D_i-1)A^{R}_{i\s})
 -k_{i+1}(D_{v_{i+1}}-1)^{-1}(\partial_{\s}g_{i+1}g_{i+1}^{-1}+g_{i+1}^{-1}\partial_{\s}g_{i+1}\\
 &-(1-D_{i+1}^T)A^L_{i+1\s}+(D_{i+1}-1)A^{R}_{i+1\s}))\,.
 \end{align}
 \begin{align}\nonumber
 \int\delta \omega|_{\partial D}=-\frac{1}{4\pi}\sum_{i=1}^{2}\int_{\partial\S}&\Tr(\delta h_ih_i^{-1},k_i(1-D^T_{v_i})^{-1}(\partial_{\tau}g_ig_i^{-1}+g_i^{-1}\partial_{\tau}g_i+(1-D_i^T)A^L_{i\tau}\\\nonumber
 &+(D_i-1)A^{R}_{i\tau})
 -k_{i+1}(D_{v_{i+1}}-1)^{-1}(\partial_{\tau}g_{i+1}g_{i+1}^{-1}+g_{i+1}^{-1}\partial_{\tau}g_{i+1}\\
 &+(1-D_{i+1}^T)A^L_{i+1\tau}+(D_{i+1}-1)A^{R}_{i+1\tau}))\,.
 \end{align}
 \begin{align}\nonumber
 \delta S|_{\partial\S}=-\frac{1}{4\pi}\sum_{i=1}^{2}\int_{\partial\S}&\Tr(\delta h_ih_i^{-1},k_i(1-D^T_{v_i})^{-1}(\partial_{+}g_ig_i^{-1}+g_i^{-1}\partial_{-}g_i+(1-D_i^T)A_{i-}\\\nonumber
 &+(D_i-1)A_{i+})
 -k_{i+1}(D_{v_{i+1}}-1)^{-1}(\partial_{+}g_{i+1}g_{i+1}^{-1}+g_{i+1}^{-1}\partial_{-}g_{i+1}\\
 &+(1-D_{i+1}^T)A_{i+1-}+(D_{i+1}-1)A_{i+1+}))\,.
 \end{align}
   
\end{document}